\documentclass[aps,prd,onecolumn,amsmath,nofootinbib,superscriptaddress,notitlepage]{revtex4-1}

\usepackage{graphicx}
\usepackage{float}
\usepackage{epstopdf}
\usepackage{epsfig}
\usepackage{amsmath}
\usepackage{color}
\usepackage{amssymb}   % \gtrsim, \geqslant, etc etc: see amsguide.ps
\usepackage{amsfonts}  % \mathfrak and \mathbb{x} (Blackboard bold)
\usepackage{amsbsy}    % \pmb and \boldsymbol
\usepackage{pstricks}
\usepackage{slashed}
\usepackage{dcolumn}
\usepackage{xparse}
\usepackage{bm}
\usepackage{hyperref}
\usepackage{soul}
\usepackage{bbm}
\usepackage[caption=false]{subfig}
\usepackage{diagbox}
\newcommand\arx[3][r]{
  \ifx r#1 \href{https://arxiv.org/abs/#2}{[arXiv:#2]} \else
  \ifx o#1 \href{https://arxiv.org/abs/#3/#2}{[arXiv:#3/#2]} \else
  \ifx b#1 \href{https://arxiv.org/abs/#2}{[arXiv:#2 [#3]]} \else
  {Illegal~option}
  \fi\fi\fi
}
\newcommand{\dd}{\mathrm{d}}

\newcommand{\blau}[1]{\textcolor{black}{#1}}

\begin{document}

%\captionsetup{justification=raggedright}

\title{Masses of compact (neutron) stars with distinguished cores}

\author{Rico Z\"ollner}\email{rico.zoellner@tu-dresden.de}
\affiliation{Institut f\"ur Technische Logistik und Arbeitssysteme, TU~Dresden, 01062 Dresden, Germany}
\author{Minghui Ding}\email{m.ding@hzdr.de}
\affiliation{Helmholtz-Zentrum Dresden-Rossendorf, 01314 Dresden, Germany}
\author{Burkhard K\"ampfer}\email{kaempfer@hzdr.de}
\affiliation{Helmholtz-Zentrum Dresden-Rossendorf, 01314 Dresden, Germany}
\affiliation{Institut f\"ur Theoretische Physik, TU~Dresden, 01062 Dresden, Germany}

%
%\date{\today}
%
\begin{abstract}
The impact of the core mass on the compact/neutron-star mass-radius relation is studied. Besides the mass, the core is parameterized by its radius and surface pressure, which supports the outside one-component Standard Model (SM) matter. The core may accommodate SM matter with unspecified (or poorly known) equation-of-state or several components, e.g.\ consisting of admixtures of Dark Matter and/or Mirror World matter etc.\ beyond the SM. Thus, the admissible range of masses and radii of compact stars can be considerably extended.   
\end{abstract}

\maketitle

%%%%%%%%%%%%%%%%%%%%%%%%%%%%%%%%%%%%%%%%%%
\section{Introduction} \label{introduction}

Strong interaction rules a variety of systems, ranging from hadrons to nuclei up to neutron stars.
The related mass scales are typically $\ge m_\pi = 0.13\, \mbox{GeV}$ for mesons,
$\ge m_p = 0.938\, \mbox{GeV}$ for baryons, $(1 \cdots 250) m_p$ for nuclei and
${\cal O} (10^{57}) m_p$ for neutron stars. ($m_{\pi, p}$ stand for pion and proton masses.)
Further systems awaiting their confirmation are
glueballs (${\cal O}(\mbox{GeV})$) and non-baryon stars, such as pion stars \cite{Brandt:2018bwq}.
Besides weak interaction, it is the long-range Coulomb interaction which limits the size (or baryon number) of nuclei,
and gravity is the binding force of matter in neutron stars. 
The phenomenon of hadron mass emergence is a fundamental issue
tightly related to non-perturbative effects in the realm of QCD, 
cf.\ \cite{Roberts:2021nhw,Roberts:2020udq} and citations therein.
Once the masses and interactions among 
hadrons are understood, one can make the journey to address masses and binding energies of nuclei
and then jump to constituents of neutron stars. Despite the notion, cool neutron stars accommodate, in the crust,
various nuclei immersed in a degenerate electron-muon environment (maybe as ``pasta'' or ``spaghetti'' or
crystalline medium). In the deeper interior, above the neutron drip density, neutrons and light clusters begin
to dominate the matter composition. These constituents and their interactions govern the mass 
(or energy density) of the medium.
At nuclear saturation density, $n_0 \approx 0.15 \, \mbox{fm}^{-3}$, one meets conditions similar to the interior
of heavy nuclei but with crucial impact of the symmetry energy when extrapolating from nuclear matter with
comparable proton and neutron numbers to a very asymmetric proton-neutron mixture. Above saturation density,
various effects hamper a reliable computation of properties of the strong-interaction medium:
three-body interactions may become even more important than at $n_0$, and
further baryon species become excited, e.g.\ strangeness is lifted from vacuum into baryons forming hyperons
whose interaction could be a miracle w.r.t.\ the hyperon puzzle \cite{Tolos:2020aln}, 
and the relevant degrees of freedom
become relativistic. Eventually, at asymptotically high density, the strong-interaction medium is converted into
quarks and gluons; color-flavor locking and color superconductivity can determine essentially the medium's
properties. The turn of massive hadronic degrees of freedom into quark-gluon excitations
is thereby particularly challenging.

While lattice QCD represents, in principle, an {\sl ab initio} approach to strong-interaction systems in all their facets,
the ``sign problem'' prevents the access to non-zero baryon number systems. Thus, the exploration of compact/neutron
stars, in particular their possible mass range, by stand-alone theory is presently not feasible.
Instead, an intimate connection of astrophysical data and compact-star modeling is required. 

%\subsection{Astrophysical data}

The advent of detecting gravitational waves from merging neutron stars,
the related multimessenger astrophysics 
\cite{Pang:2021jta,Annala:2021gom,Yu:2021nvx,Nicholl:2021rcr,Margutti:2020xbo,Tang:2020koz,Tews:2020ylw,Silva:2020acr}
and the improving mass-radius determinations
of neutron stars, in particular by NICER data \cite{Riley:2021pdl,Miller:2021qha,Miller:2019cac,Riley:2019yda,Raaijmakers:2021uju}, 
stimulated a wealth of activities. Besides masses and radii, moments of inertia
and tidal deformabilities become experimentally accessible and can be confronted with theoretical models 
\cite{Chatziioannou:2020pqz,Christian:2019qer,Motta:2022nlj,Jokela:2021vwy,Kovensky:2021kzl,Zhang:2021xdt,Pereira:2020cmv}.
The baseline of the latter ones is provided by non-rotating, spherically symmetric
cold dense matter configurations. The sequence of white dwarfs (first island of stability)
and neutron stars (second island of stability) and --possibly \cite{Christian:2020xwz}-- 
a third island of stability \cite{Gerlach:1968zz,Kampfer:1981zmq,Kampfer:1981yr,Haensel:1987,Haensel:1983}
shows up thereby when going to more compact objects, with details depending sensitively
on the actual equation of state (EoS). 
The quest for a fourth island has been addressed too \cite{Li:2019fqe,Alford:2017qgh}.
``Stability'' means here the damping of radial disturbances, at least. Since the radii of configurations
of the second (neutron stars) and third (hypothetical quark/hybrid stars 
\cite{Malfatti:2020onm,Pereira:2022stw,Bejger:2016emu,Li:2021sxb,Cierniak:2020eyh,Ranea-Sandoval:2015ldr,Alford:2015gna,Tan:2021ahl}) 
islands are very similar, 
the notion of twin stars \cite{Glendenning:1998ag,Jakobus:2020nxw,Li:2019fqe,Alford:2017qgh,Malfatti:2020onm}
has been coined for equal-mass configurations; ``masquerade'' was another related term \cite{Alford:2004pf}.

%\subsection{Relation to heavy-ion collisions}

We emphasize the relation of ultra-relativistic heavy-ion collision physics, probing
the EoS $p(T, \mu_B \approx 0)$, and compact star physics, probing
$p(T \approx 0, \mu_B)$ when focusing on static compact-star properties
\cite{Klahn:2006ir,Most:2022wgo}.
(Of course, in binary or ternary compact-star merging-events, 
also finite temperatures $T$ and a large range of baryon-chemical potential $\mu_B$ are probed 
which are accessible in medium-energy heavy-ion collisions \cite{HADES:2019auv}.)
Implications of the conjecture of a first-order phase transition at small temperatures and 
large baryo-chemical potentials or densities
\cite{Stephanov:1999zu,Karsch:2001cy,Fukushima:2010bq,Halasz:1998qr}
can also be  studied by neutron-hybrid-quark stars \cite{Blaschke:2020vuy,Blacker:2020nlq,Cierniak:2021knt,Orsaria:2019ftf}.
It is known since some time 
\cite{Gerlach:1968zz,Kampfer:1985mre,Kampfer:1983we,Kampfer:1981yr,Kampfer:1981zmq}
that a cold EoS with special pressure-energy density relation $p(e)$, e.g.\ a strong local softening
up to first-order phase transition with a density jump, 
can give rise to a ``third family'' of
compact stars, beyond white dwarfs and neutron stars. In special cases, the third-family stars appear
as twins of neutron stars \cite{Schertler:2000xq,Alford:2004pf,Christian:2017jni}.
Various scenarios of the transition dynamics to the denser configuration as
mini-supernova have been discussed also quite early \cite{Migdal:1979je,Kampfer:1983zz}. 

%\subsection{Beyond Standard Model}

While the Standard Model (SM) of particle physics seems to accommodate nearly all of the observed
phenomena of the micro-world, severe issues remain. Among them is the $(g-2)_\mu$ puzzle or
the proton's charge radius. Another fundamental problem is the very nature of Dark Matter (DM): 
Astrophysical and cosmological observations seem to require inevitably its existence, 
but details remain elusive despite many concerted attempts, e.g.\ 
\cite{DiLuzio:2021gos,Bertone:2004pz,Tulin:2017ara,Hodges:1993yb}.
Supposed DM behaves like massive particles,
these could be captured gravitationally in the centers of compact stars 
\cite{Karkevandi:2021ygv,Dengler:2021qcq,Hippert:2022snq},
thus providing a non-SM component there.
This would be an uncertainty on top of the less reliably known SM-matter state.
Beyond the SM, also other feebly interacting particles could populate compact stars. A candidate scenario is
provided, for instance, by Mirror World \blau{(MW)}
\cite{Alizzi:2021vyc,Goldman:2019dbq,Beradze:2019yyp,Berezhiani:2021src,Berezhiani:2020zck}
i.e.\ a parity-symmetric complement to our SM-world with very tiny
non-gravity interaction. There are many proposals of portals from our SM-world to such beyond-SM scenarios,
cf.\ \cite{Beacham:2019nyx}.

%\subsection{Outline of the paper}

Guided by these remarks we follow here an access to static cold compact stars already launched in \cite{Zollner:2022dst}:
We describe the core by a minimum of  parameters and determine the resulting compact-star masses and radii 
by assuming the knowledge of the equation of state of the SM-matter enveloping the core. 
A motivation is the quest of a mass gap between compact stars and black holes.

Our paper is organized as follows. In Section \ref{sect:II} we recall the \blau{Tolman-Oppenheimer-Volkoff} equations, their scaling property
and introduce our core-corona decomposition.
Small cores with and without MW/DM admixtures are considered in Section \ref{sect:small_core}.
Section \ref{sect:III} is devoted to the core-corona decomposition, where a specific EoS is deployed
for the explicit construction. We summarize in Section \ref{sect:summary}. 
We supplement our paper in Appendix \ref{sect:A} by a brief retreat to the emergence of hadron masses \blau{as key issue in understanding typical scales of compact (neutron) star masses.}
Appendix \ref{sect:holography} sketches \blau{a complementary approach:} the construction of an EoS by holographic means, thus
transporting information of a hot \blau{(quark-gluon)} QCD EoS to a cool EoS
and connecting heavy-ion collisions and compact star physics. \blau{These appendices survey (Appendix~\ref{sect:A}) and exemplify (Appendix~\ref{sect:holography}) symmetry and governing principles where the access to astrophysical objects is based upon.}
%%%%%%%%%%%%%%%%%%%%%%%%%%%%%%%%%%%%%%%%%%

\section{One-component static cool compact stars: TOV equations} \label{sect:II}

The standard modeling of compact star configurations is based on 
the Tolman-Oppen-heimer-Volkoff (TOV) equations
\begin{align}
\frac{\mbox{d} p}{\mbox{d} r } &= 
- G_N \frac{[e(r)+p(r)] [m(r)  + 4 \pi r^3 p(r)]}{r^2 [1 - \frac{2 G_N m(r)}{r}]}, \label{eq:p_prime} \\
\frac{\mbox{d} m}{\mbox{d} r} &= 4 \pi r^2 e(r), \label{eq:m_prime}
\end{align}  
resulting from the energy-momentum tensor of a one-component static isotropic fluid 
(described locally by pressure $p$ and energy density $e$ as only quantities relevant for the medium) 
and spherical symmetry of both space-time and matter, 
within the framework of Einstein gravity without cosmological term \cite{Schaffner-Bielich:2020psc}.
Newton's constant is denoted by $G_N$, and
natural units with $c = 1$ are used,
unless when relating mass and length and energy density, where $\hbar c$ is needed.

\subsection{Scaling of TOV equations and compact/neutron star masses and radii} \label{sect:2-1}

The TOV equations become free of any dimension by the scalings
\begin{align} \label{eq:scaling}
e =\mathfrak{s} \bar e, \quad 
p = \mathfrak{s} \bar p, \quad 
r = (G_N \mathfrak{s})^{-1/2} \bar r, 
\quad m =  (G_N^3 \mathfrak{s})^{-1/2} \bar m,
\end{align}
where $\mathfrak{s}$ is a mass dimension-four quantity or has dimension of energy density.
It may be a critical pressure of a phase transition or a limiting energy density, e.g.\ at the boundary matter-vacuum 
\cite{Schulze:2009dy}.
Splitting up $\mathfrak{s} = n m_p$ into a number density $n$ and the energy scale $m_p$, one gets
$(G_N \mathfrak{s} )^{-1/2} = 7~\mbox{km} / \sqrt{10^{-2} n / n_0}$ and
$(G_N^3 \mathfrak{s} )^{-1/2} = 4.8  M_\odot / \sqrt{10^{-2} n / n_0}$,
where the nuclear saturation density $n_0 = 0.15$~fm${}^{-3}$ is used as reference density.
The scales $m_{\pi, p}$ facilitate the densities $n = m_\pi^3 \to 2.3 n_0$ and $n = m_p^3 \to 833 n_0$. 
That is, the scale solely set by the nucleon mass, i.e.\ $\mathfrak{s} = m_p^4$ \cite{Baym:2017whm},
%\marginpar{\color{red} {\footnotesize check! \cite{Baym:2017whm} quotes \underline{17.2~km}, i.e. factor $2 \pi$ larger}}
yields $(G_N \mathfrak{s})^{-1/2} = 2.74~\mbox{km}$ and
$(G_N^3 \mathfrak{s} )^{-1/2} = 1.86  M_\odot$, suggesting that the nucleon mass 
(see Appendix~A on its emergence in QCD) determines gross properties of neutron stars, such
as mass and radius (modulus factor $2 \pi$)
in an order-of magnitude estimate.\footnote{
In contrast, the density estimate via $M = \frac{4 \pi}{3}  \langle n \rangle m_p R^3$
yields, independently of $G_N$,  
$\langle n \rangle = 2 n_0 \frac{M}{M_\odot} \frac{10^3}{(R/\mbox{km})^3} \approx 4 n_0$
for $M = 2 M_\odot$ and $R = 10$~km, thus pointing to the importance of 
actual numerical values of $\bar r$ and $\bar m$.} 

In fact, recent measurements and supplementary work report averaged mean and
individual heavy compact/neutron stars masses and radii  (often on 67\% credible level) as follows 

\begin{center}
\renewcommand{\arraystretch}{1.5} % Default value: 1
\begin{tabular}{ l | l l  } 
PSR & $M~[M_\odot]$ & $R$~[km] \\ 
\hline
       & 1.4 & $11.94^{ + 0.76} _{- 0.87}$ ${}^{1)}$ \cite{Pang:2021jta},  %%%\\
$12.45 \pm 0.65$ \cite{Miller:2021qha},
%          &       & 
$12.33 ^{+ 0.76} _{- 0.81}$ \cite{Riley:2021pdl}\\
J0030+0451 &  $1.34 ^{+ 0.15} _{- 0.16}$ & $12.71 ^{+ 1.14} _{- 1.19}$ \cite{Riley:2019yda}\\
                   & $1.44 ^{+ 0.15} _{- 0.14}$ &  $13.02 ^{+ 1.24} _{- 1.06}$ \cite{Miller:2019cac}, 
%                   &                                 & 
$12.18 ^{+ 0.56} _{- 0.79}$ \cite{Raaijmakers:2021uju}\\
J1614–2230 & $1.908 \pm 0.016 $ \cite{Demorest:2010bx} & \\
J0348+0432 & $2.01 \pm 0.04 $ \cite{Antoniadis:2013pzd} &  \\
J0740+6620 & $2.072 ^{+ 0.067} _{- 0.066}$ & $12.39 ^{+ 1.30} _{- 0.98}$ \cite{Riley:2021pdl} \\    
                   & $2.08 \pm 0.07 $  \cite{Fonseca:2021wxt} & $13.7 ^{+ 2.6} _{- 1.5}$${}^{2)}$ \cite{Miller:2021qha},
%&                   &                                                               &
$11.96 ^{+ 0.86} _{- 0.81}$  \cite{Pang:2021jta} \\
0952-0607${}^{3)}$   & $2.35 \pm 0.17$ \cite{Romani:2022jhd} &  \\      
\end{tabular}
\end{center}
{\footnotesize
${}^{1)}$90\% confidence.
${}^{2)}$With nuclear physics constraints at low density and gravitational radiation data from GW170817 added in, 
the inferred radius drops to $(12.35 \pm 0.75)$ km \cite{Miller:2021qha}. 
${}^{3)}$Black-widow binary pulsar PSR~0952-0607. }\\[1mm]

One has to add GW190814: gravitational waves from the coalescence of a 23 solar mass \blau{black hole} 
with a $2.6 M_\odot$  compact object \cite{LIGOScientific:2020zkf}.
An intriguing question concerns a possible mass gap between compact-star maximum-mass 
\cite{Ecker:2022dlg} and light black holes, cf.\  \cite{Shao:2022qws,Farah:2021qom}.

\subsection{Solving TOV equations}

Given a unique relationship of pressure $p$ and energy density $e$ as EoS
$e(p)$, in particular at zero temperature, the TOV equations are integrated customarily with boundary conditions
$p(r = 0) = p_c $ and $m(r = 0) = 0$
(implying $p(r) = p_c - {\cal O} (r^2)$ and $m(r) = 0 + {\cal O} (r^3)$ at small radii $r$),
and $p(R) = 0$ and $m(R) = M$ with $R$ as circumferential radius and $M$ as gravitational mass
(acting as parameter in the external (vacuum) Schwarzschild solution at $r > R$).
The quantity $p_c$ is the central pressure.
The solutions $R(p_c)$ and $M(p_c)$
provide the mass-radius relation \blau{$M(R)$} in parametric form.

A big deal of efforts is presently concerned about the EoS at supra-nuclear densities \cite{Reed:2021nqk}.
For instance,
Fig\blau{.}~1 in \cite{Annala:2019puf} exhibits the currently admitted uncertainty: up to a factor of ten in pressure 
as a function of energy density. At asymptotically large energy density, perturbative QCD constraints
the EoS, though it is just the non-asymptotic supra-nuclear density region which determines crucially
the maximum mass and
whether twin stars may exist or quark-matter cores appear in neutron stars. 
Accordingly, one can fill this gap by a big number (e.g.\ millions \cite{Altiparmak:2022bke})
of test EoSs to scan through the possibly resulting manifold of mass-radius curves, see 
\cite{Ayriyan:2021prr,Greif:2020pju,Lattimer:2015nhk,LopeOter:2019pcq}.
However, the possibility that neutron stars may accommodate other components than
Standard Model matter, e.g.\ exotic material as Dark Matter
\cite{Anzuini:2021lnv,Bell:2019pyc,Das:2021hnk,Das:2021yny},
can be an obstacle for the safe theoretical modeling of a concise mass-radius relation in such a manner. Of course, 
inverting the posed problem with sufficiently precise data of masses and radii as input offers 
a promising avenue towards determining the EoS 
\cite{Newton:2021yru,Huth:2021bsp,Ayriyan:2021prr,Blaschke:2020qqj,Raaijmakers:2021uju,Raaijmakers:2019dks,Raaijmakers:2019qny}.

Here, we pursue another perspective. We parameterize the supra-nuclear core by a radius~$r_x$ and the
included mass~$m_x$ and integrate the above TOV equations only within the corona,\footnote{Our notion 
``corona'' is a synonym for  ``mantel'' or ``crust'' or ``envelope'' or ``shell''.
It refers to the complete part of the compact star outside the core, $r_x \le r \le R$.} 
i.e.\ from pressure $p_x$
to the surface, where $p = 0$. This yields the total mass $M(r_x, m_x; p_x)$ and the total radius $R(r_x, m_x; p_x)$
by assuming that the corona EoS $e(p)$ is reliably known at $p \le p_x$
and only SM matter occupies that region. Clearly, without knowledge
of the matter composition at $p > p_x$ (may it be SM matter with an uncertainly known EoS or
may it contain a Dark-Matter admixture, for instance, or monopoles or some other type of
``exotic'' matter) one does not get a simple mass-radius relation by such a procedure, 
but admissible area(s) over the mass-radius plane, depending on the core
parameters $r_x$ and $m_x$ and the matching pressure $p_x$ and related energy density $e_x$.
This is the price of avoiding a special model of the core matter composition.

If the core is occupied by a one-component SM medium, the region $p > p_x$ and $e > e_x$ can be mapped out by many test EoSs which obey locally the constraint $v_s^2 \in [0,1]$ to obtain the corresponding region in the mass-radius plane, cf.\  Fig.~2 in \cite{Gorda:2022jvk} for an example processed by Bayesian inference.
This is equivalent, to some extent, to our core-corona decomposition for SM matter-only.

\section{Small-core approximation and beyond}\label{sect:small_core}

\subsection{One-component core}

For small one-component distinguished cores one can utilize the EoS parameterization
from a truncated Taylor expansion of $p(e) \ge p_x$ at $\lambda e_x$,
$p(e) = p(\lambda e_x) + \frac{\partial p}{\partial e} \vert_{\lambda e_x} (e - \lambda e_x) + \cdots$,
\begin{align} \label{eq:coreEoS}
p(e) = p_x + v_s^2 (e - \lambda e_x),
\end{align} 
where $\lambda > 1$ is for a density jump at the core boundary
and $\lambda = 1$ continues continuously (but may be kinky) the corona EoS at $p \ge p_x$;
$p_x$ and $e_x$ mark the ``end point'' of the corona EoS. 
A Taylor expansion for small cores,
\begin{align}
p (r) &= \sum_{i=0}^\infty p_{2 i} r^{2 i}, \\
m(r) &=  \sum_{i=0}^\infty m_{2 i + 1} r^{2 i +1},
\end{align}
gives by means of Eqs.~(\ref{eq:p_prime}, \ref{eq:m_prime}) with (\ref{eq:coreEoS})\footnote{
That is $e = \lambda e_x - v_s^{-2} p_x  + v_s^{-2} \sum_{i=0}^\infty p_{2 i} r^{2 i}$.
}
and $p_{2 i + 1} = 0$ and $m_{2 i} = 0$ for all $i \in \mathbb{N}_0$
\begin{align}
m_1 = 0, \,
%m_3 = \frac{4 \pi}{3} [\lambda e_x - v_s^{-2} p_x] + 4 \pi v_s^{-2} p_0, \quad
m_3 = \frac{4 \pi}{3} e_c, \,
m_{2 i + 1} = \frac{4 \pi v_s^{-2}}{2 i + 1} p_{2 i -2}  \, \mbox{for} \, i \ge 2 ,
\end{align}
where $e_c  \equiv e(p_c) =  \lambda e_x + v_s^{-2} (p_c - p_x)$,
and the recurrence 
\begin{align}
p_0 &= p_c, \quad
p_2 = - \frac{2 \pi}{3} G_N (e_c + p_c)(e_c + 3 p_c), \\
p_{2 i} &= \frac{G_N}{2i} \left(2 A_i -[\lambda e_x - v_s^{-2} p_x](m_{2 i +1} + 4 \pi p_{2 i -2}) - (1+v_s^{-2}) B_{i - 1} \right) \, \mbox{for} \, i \ge 1 , \label{eq:pi_recurr} \\
& A_i = \sum_{j=0}^i 2(i-j) m_{2j} \, p_{2i-2j}, \tag{\ref{eq:pi_recurr}'} \\
& B_i = \sum_{j=0}^i (m_{2j+3} + 4 \pi p_{2j}) \, p_{2i-2j} . \tag{\ref{eq:pi_recurr}''}
\end{align}
In leading order one obtains
\begin{align}
r_x &\approx \frac{ \delta^{1/2} (1 - {\cal O}(\delta)) }{ \sqrt{\frac{2 \pi}{3} G_N p_x W } }
\approx 
\frac{60.1~\mbox{km}}{\sqrt{W} } 
\sqrt{\delta \frac{100~\mbox{MeV/fm}^3}{p_x}}  , \label{eq:rx_approx}\\
m_x &\approx 2 \lambda \frac{e_x}{p_x} 
\frac{ \delta^{3/2} (1 - {\cal O}(\delta))}{ \sqrt{\frac{2 \pi}{3} G_N^3 p_x W^3  }  }
\approx
\frac{81.2 M_\odot}{W^{3/2}}  \lambda \frac{e_x}{100~\mbox{MeV/fm}^3} \times \left( \delta \frac{100~\mbox{MeV/fm}^3}{p_x} \right)^{3/2} , \label{eq:mx_approx}
\end{align}
where $\delta := p_c / p_x - 1$ and
$W := 3 + 4 \lambda \frac{e_x}{p_x} + \lambda^2 \left( \frac{e_x}{p_x} \right)^2 $.
The scale setting is by $p_x$ and $\lambda e_x / p_x$. The core-mass--core-radius relation is
$m_x (\delta) \approx \frac{4 \pi}{3} p_x \lambda \left(\frac{e_x}{p_x}\right) r_x( \delta)^3 (1 + {\cal O}(\delta))$.

\begin{figure}[t!]
\includegraphics[width=0.49\columnwidth]{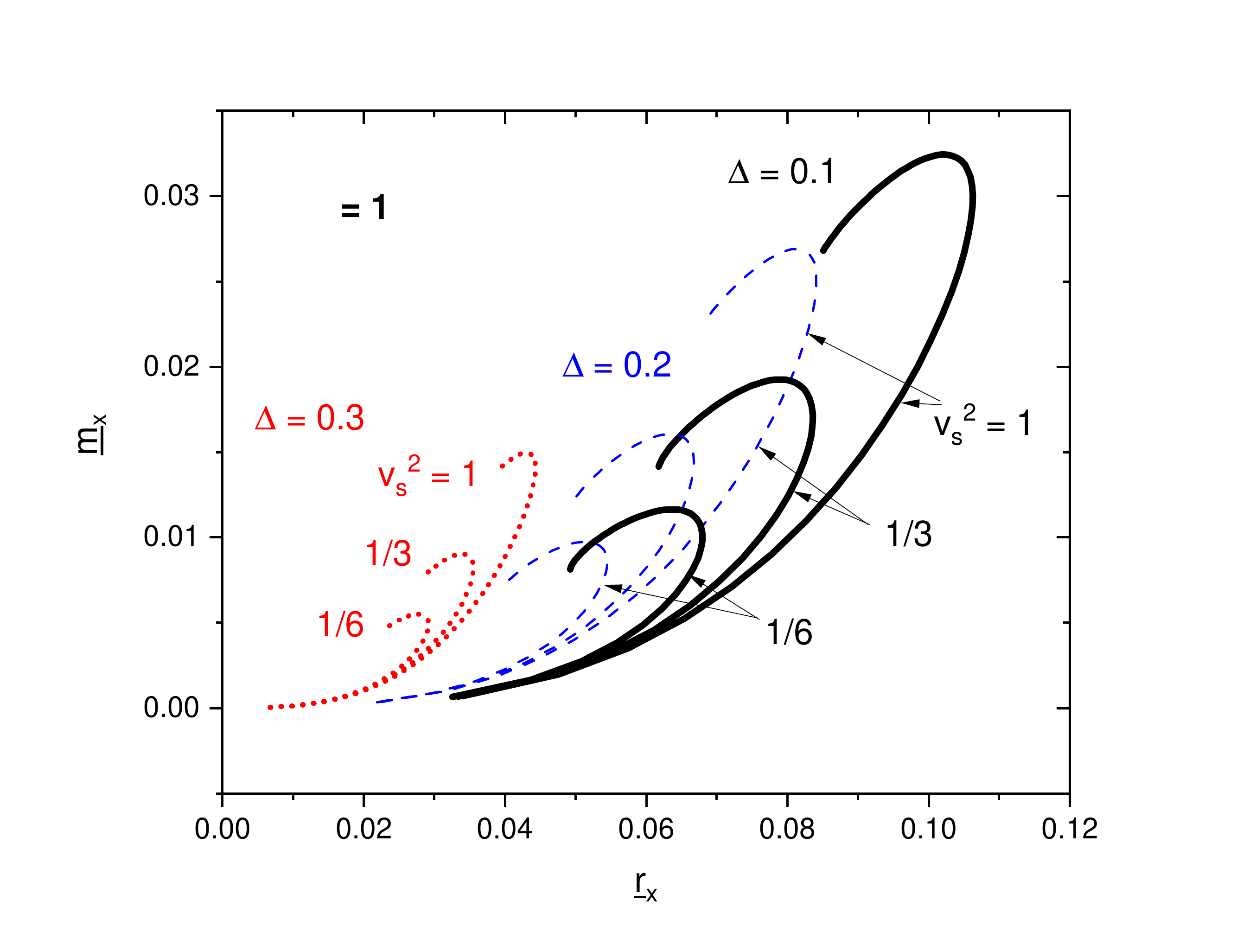}   %\hspace*{-9mm}
\includegraphics[width=0.49\columnwidth]{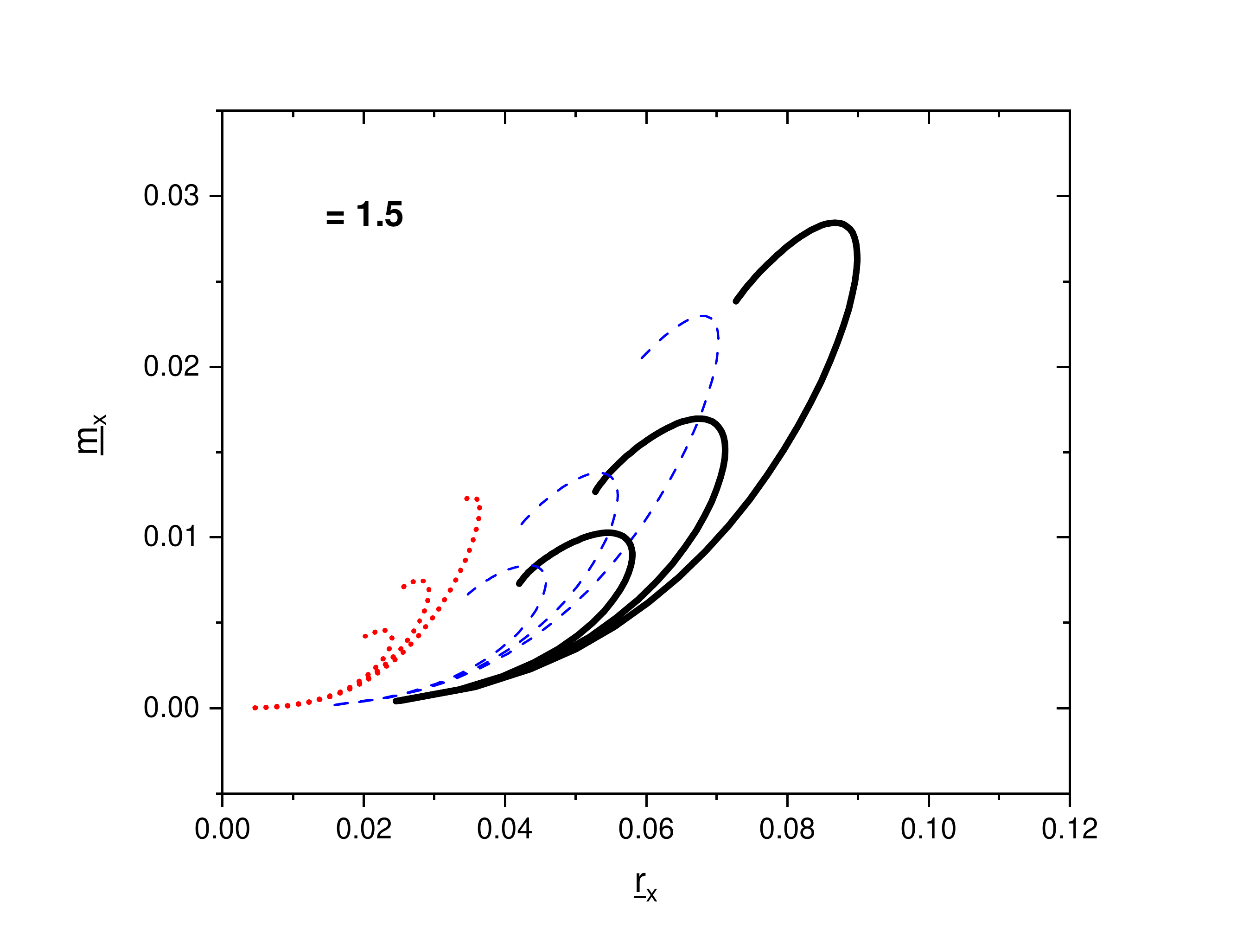}
\caption{Scaled core masses $\bar m_x$ as a function of the scaled core radii $\bar r_x$ for various values of
$\Delta^{corona}=\Delta$ and $v_s^2$ for a one-component medium with EoS (\ref{eq:coreEoS}).
Left (right) 
panel is for $\lambda = 1$ (1.5). 
The scaling quantity is $\mathfrak{s} = p_x$.
For central pressures $\bar p_c = 1.1^n$, $n= 1 \cdots 55$.
The displayed curves are limited by $\bar m_x < \bar r_x / 2$ (black hole limit)
and $\bar m_x > \frac{4 \pi}{3} \lambda \frac{3}{1 - 3 \Delta^{corona}} \bar r_x^3$.
The latter expression is for the respective asymptotic curve in the small-$\bar r_x$ region. 
To convert to usual dimensions one employs 
$r_x = \bar r_x \frac{86.9~\mbox{km}}{ \sqrt{p_x / 100~\mbox{[MeV/fm}^3]} }$ and
%$r_x = 86.9~\mbox{km} \frac{\bar r_x}{ \sqrt{p_x / 100~\mbox[MeV/fm}^3]} }$ and
$m_x = \bar m_x \frac{58.8~M_\odot}{ \sqrt{p_x / 100~\mbox{[MeV/fm}^3]} }$, where 
%$m_x = 58.9~M_\odot \frac{\bar m_x}{ \sqrt{p_x / 100~\mbox{[MeV/fm}^3]} }$, where
``$p_x / 100~\mbox{[MeV/fm}^3]$'' denotes the scaling pressure $p_x$ in units of 100~MeV/fm${}^3$.
The approximations (\ref{eq:rx_approx}, \ref{eq:mx_approx}) apply only in the
small-$\bar m_x$ and small-$\bar r_x$ regions.
\label{fig:small_cores} 
}
\end{figure}

To control and extend the above approximations we solve numerically the scaled TOV equations
by assuming that Eq.~(\ref{eq:coreEoS}) holds true in the core. 
(Of course, this is an {\sl ad hoc} assumption aimed at providing an explicit example of mass-radius
relations of the core.) The core-corona matching is at $p_x$, i.e.\ the maximum (minimum)
pressure of the corona (core) EoS.
The related energy density is $e_x = 3 p_x / (1 - 3 \Delta^{corona})$, where $\Delta^{corona}$ denotes
the trace anomaly measure discussed below in Eq.~(\ref{eq:Delta}). It is used here for a suitable parameterization
of the double $(e_x, p_x)$ \blau{by means of the corona EoS}. The core EoS Eq.~(\ref{eq:coreEoS}), $e(p) = \lambda e_x + v_s^{-2} (p - p_x)$,
enters, after scaling according to Eq.~(\ref{eq:scaling}), the dimensionless TOV equations
\begin{align}
\frac{\mbox{d} \bar p}{\mbox{d} \bar r } &=
- \frac{[\bar e(\bar r)+\bar p(\bar r)] [\bar m (\bar r)  + 4 \pi \bar r^3 \bar p(\bar r)]}
{\bar r^2 [1 - \frac{2 \bar m(\bar r)}{\bar r}]}, 
\tag{\ref{eq:p_prime}'}
\label{eq:p_prime_scaled} \\
\frac{\mbox{d} \bar m}{\mbox{d} \bar r} &= 4 \pi \bar r^2 \bar e(\bar r), 
\tag{\ref{eq:m_prime}'}
\label{eq:m_prime_scaled}
\end{align} 
for $\bar p \in [\bar p_c, \bar p_x]$ to get $\bar r_x (\bar p_c)$ and $\bar m_x (\bar p_c)$. The corresponding
scaled core mass vs.\ core radius relations are exhibited in Fig.~\ref{fig:small_cores}, where the
scaling quantity $\mathfrak{s} = p_x$ is employed, i.e.\ $\bar p_x = 1$.
The figure offers a glimpse on the systematic of the parameter dependence. Note that it applies to
all values of $p_x > 0$. The finite pressure and energy density at core boundary facilitates a pattern
of mass-relations and dependence on sound velocity as known from bag model EoS, cf.\ Fig.~7 in \cite{Schulze:2009dy}.
$\lambda > 1$ causes an overall shrinking of the pattern plus a slight up-shift,
see right panel for $\lambda = 1.5$.
Considering, e.g., $e_x \in [150, 1500]$~MeV/fm${}^3$ and the scaling $\propto 1/\sqrt{e_x}$, cf.\ (\ref{eq:scaling}),
the core masses and radii change by a factor up to three, depending on the actual value of $e_x$.

\begin{figure}[t!]
\includegraphics[width=0.49\columnwidth]{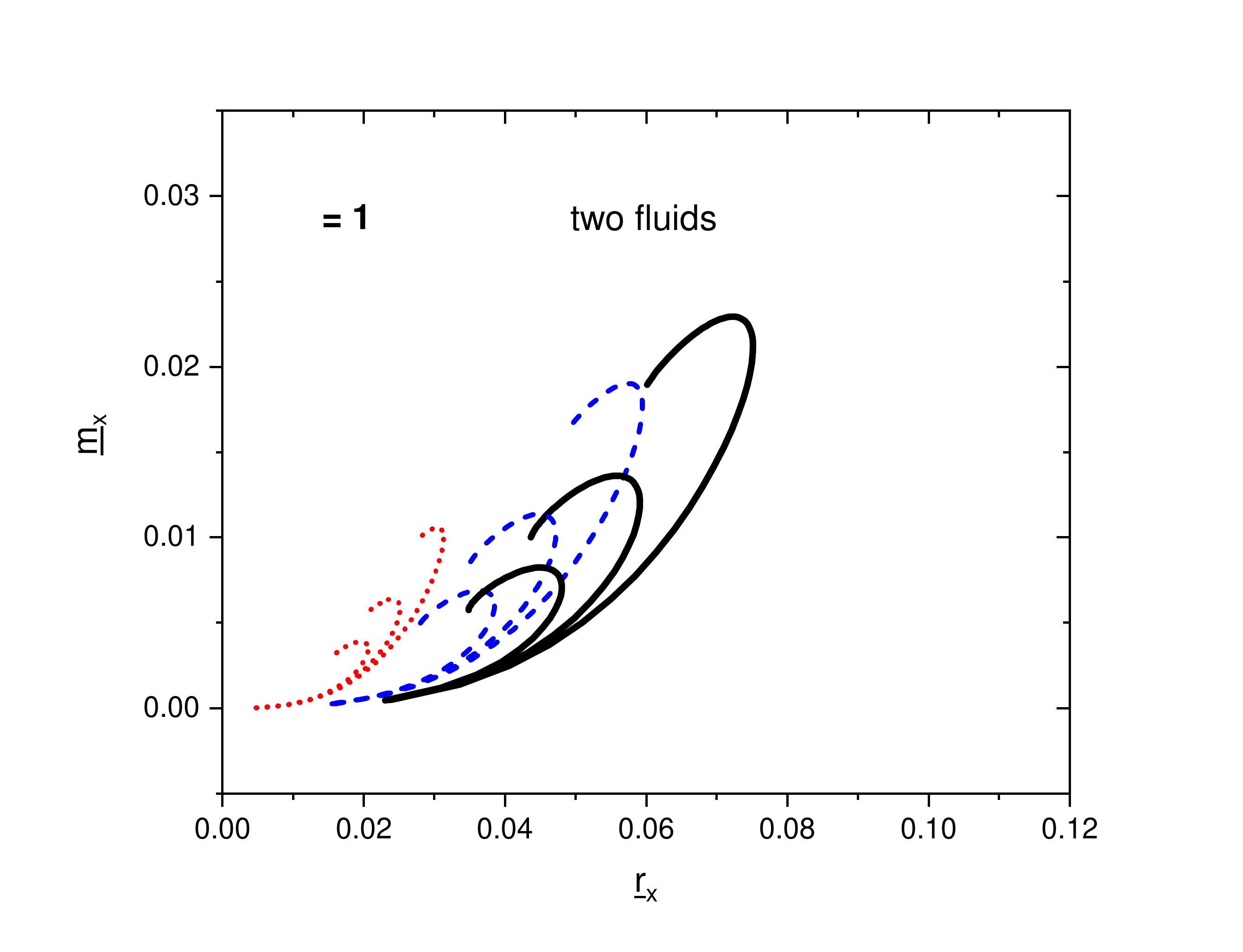}   %\hspace*{-9mm}
\includegraphics[width=0.49\columnwidth]{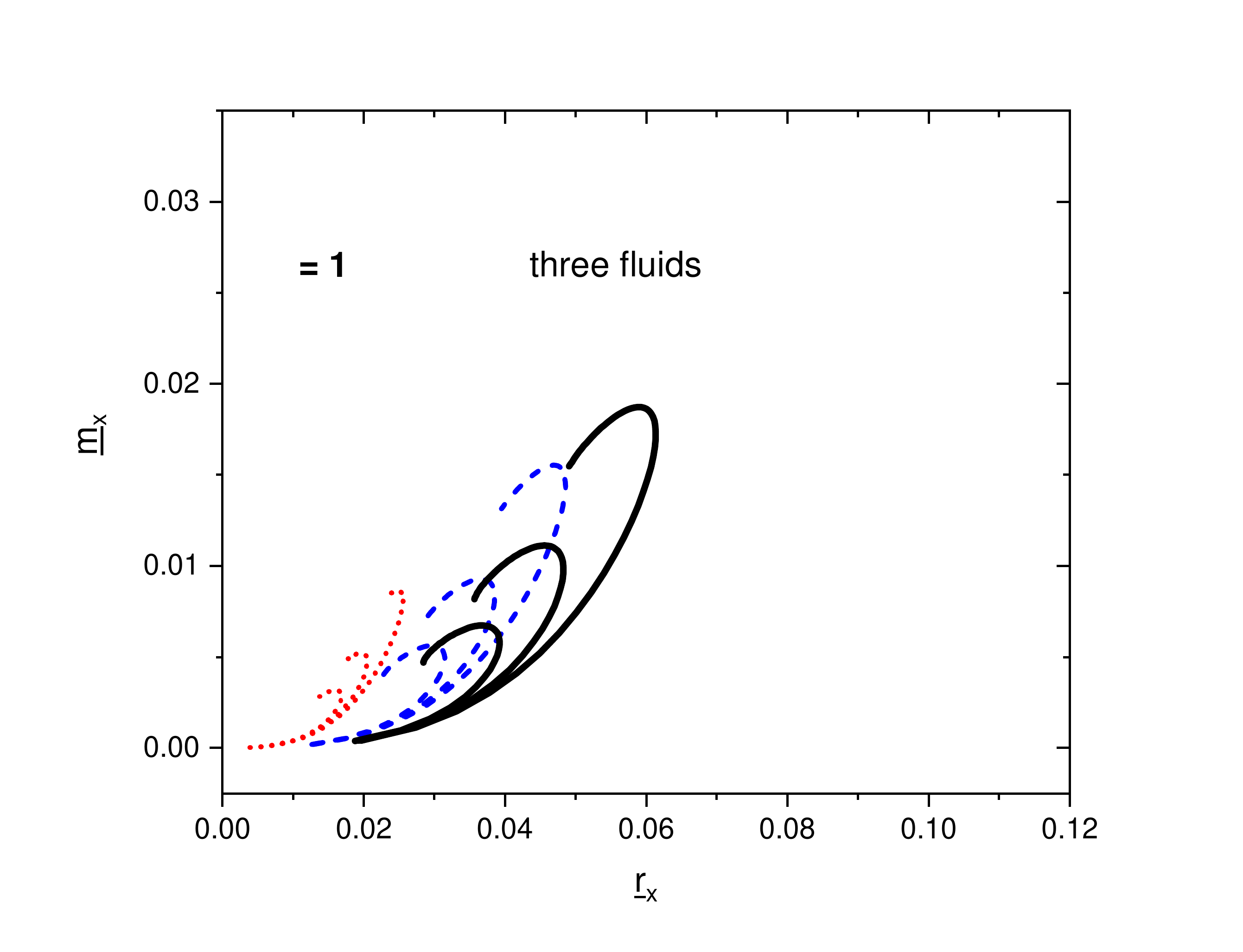}
\caption{Left panel: As left panel of Fig.~\ref{fig:small_cores} but for a two-component medium of SM + MW matter
with $p_{(1)} (r) = p_{(2)} (r)$ and EoS  (\ref{eq:coreEoS}) for both components.
Note the difference to the two-fluid core-shell construction in \cite{Cassing:2022tnn}.
Right panel: As left panel but for a three-component medium.
\label{fig:1MW} 
}
\end{figure}

\subsection{Multi-component cores}
 
In a multi-component medium\footnote{
Note that our considerations apply to components which do not mutually interact in 
a direct microscopical manner, but are coupled solely by the common gravity field.}
with known EoSs one has to add the contributions by
$ [m(r)  + 4 \pi r^3 p(r)] \to \sum_i  [m_{(i)}(r)  + 4 \pi r^3 p_{(i)}(r)]$ and
$[1 - \frac{2 G_N m(r)}{r}] \to  [1 - \frac{2 G_N \sum_i m_{(i)}(r)}{r}]$
and solve in parallel the multitude of TOV equations for the components $i$.
A particular minimum-parameter model is Mirror World matter (component $i = 2$) which is
completely symmetric to SM matter (component $i = 1$), i.e.\ $p_{(1)} (r) = p_{(2)} (r)$ etc.,
turning $\sum_{(i)} \to 2$. The results are exhibited in Fig.~\ref{fig:1MW}-left.
The increased energy density by the two superimposed fluids cause a shrinking of the core-mass--core-radius
curves similar to the increase of $\lambda$ in Fig.~\ref{fig:small_cores}.
The extension to three components is exhibited in Fig.~\ref{fig:1MW}-right.
Here, $\sum_{(i)} \to 3$ applies. Due to further increased energy density and pressure,
the configurations become even more ``compact'', i.e.\ core masses and core radii shrink further on. 

One may quantify the core compactness by
$\bar {\cal C}_x := 2 \bar m_x^{max} / \bar r_x\vert_{ \bar m_x^{max}}$ and note that,
within the scanned parameter patch, 
it (i) is independent of the number of fluids,
(ii) increases slightly with \blau{$\Delta^{corona}$} at $v_s^2 = const$ and
(iii) decreases with decreasing  $v_s^2 = const$ at \blau{$\Delta^{corona} = const$}, see Table~\ref{tab:I}. \blau{Note that the usual definition of compactness is without the factor of two.}

\begin{table}
%\begin{SCtable}[50]
\begin{center}
\begin{tabular}{ c | c c c } 
\diagbox{$v_s^2$}{$\Delta$}   &  0.1  &  0.2  &  0.3 \\
\hline
 1   & 0.64 & 0.67 & 0.70 \\
1/3 & 0.49 & 0.51 & 0.53 \\
1/6 & 0.37 & 0.38 & 0.40 \\
\end{tabular}
\end{center}
\caption{Core compactness $\bar {\cal C}_x  = 2 \bar m_x^{max} / \bar r_x\vert_{ \bar m_x^{max}}$ 
for various values of $v_s^2$ \blau{(sound velocity squared in the core)} and \blau{$\Delta^{corona}=\Delta$}. \blau{For $\lambda = 1$.}}  
\label{tab:I}
%\end{SCtable}
\end{table}

Of course, more general is an asymmetric Mirror World component facilitating
$p_{(1)} (r) \ne p_{(2)} (r)$, especially  $p_{(1)} (r = 0)  \ne p_{(2)} (r = 0)$, even
for the common EoS (\ref{eq:coreEoS}), dealt with in Appendix A in \cite{Zollner:2022dst}.
We do not dive into various conceivable scenarios here and refer the interested reader to \cite{Cassing:2022tnn,Kain:2021hpk}.

\section{Core-corona decomposition with NY$\Delta$ DD-EM2 EoS}\label{NYD}\label{sect:III}

\subsection{Trace anomaly}

\begin{figure}[t!]
\includegraphics[width=0.49\columnwidth]{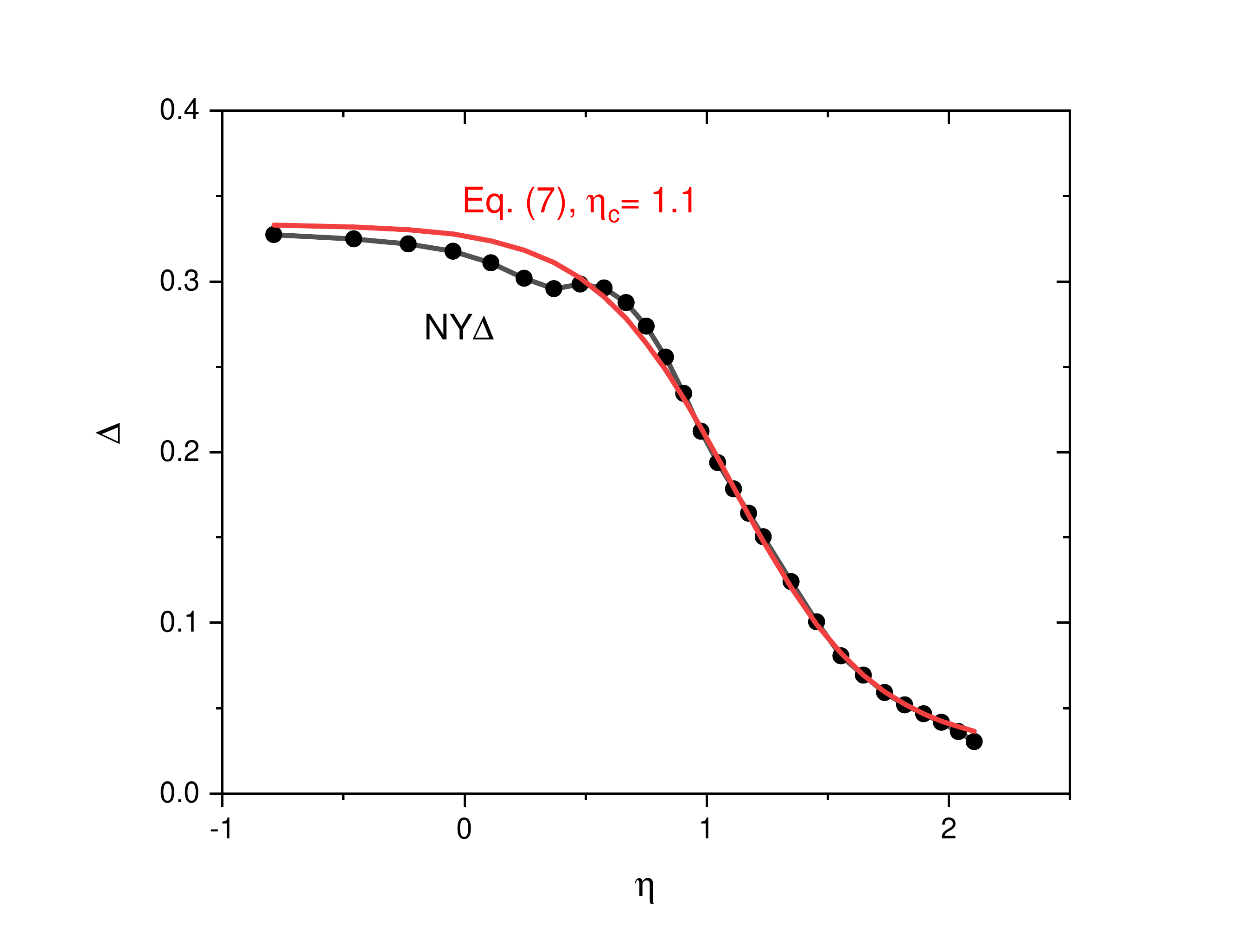}   % \hspace*{-9mm}
\includegraphics[width=0.49\columnwidth]{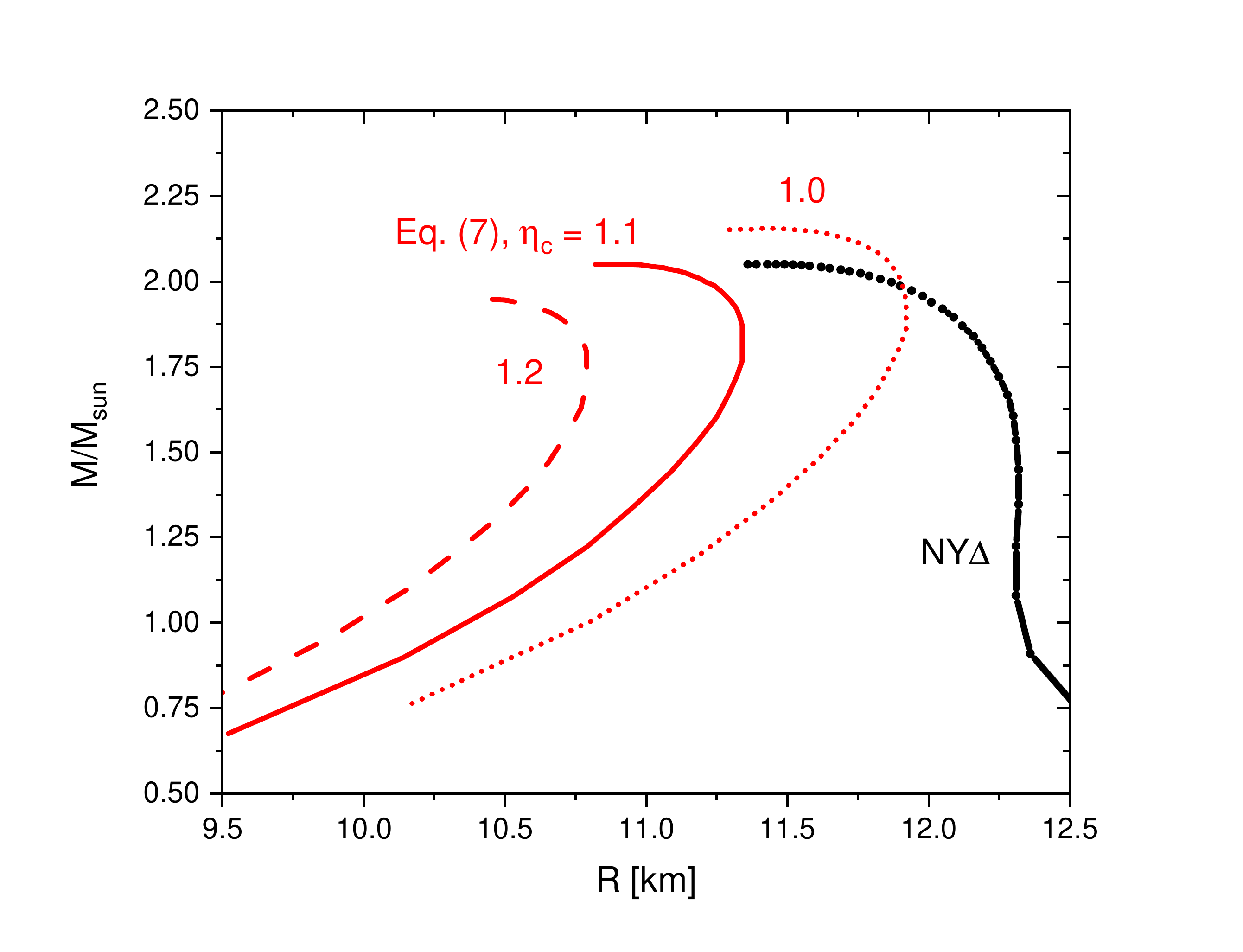}
\caption{Left panel: Trace anomaly measure $\Delta = 1/3 - p/e$ 
as a function of $\eta \equiv \ln e/(n_0 m_p)$
for the NY$\Delta$ DD-EM2 EoS of \cite{Li:2019fqe}
(black curve with symbols) and 
the adjusted parameterization by Eq.~(7) in \cite{Fujimoto:2022ohj}
(red curve; $\eta_c = 1.1$, other parameters as in \cite{Fujimoto:2022ohj}, 
see Eq.~(\ref{eq:LarryEoS}) below).
NY$\Delta$ DD-EM2 shows a softening at $p \approx 10$~MeV/fm${}^3$,
$\eta \approx 0.5$, caused by the onset of $\Delta$ proliferation,
similar to \cite{Marczenko:2022dne}.
Right panel: Mass-radius relation for NY$\Delta$ DD-EM2 EoS of \cite{Li:2019fqe} (solid black curve)
in comparison with the parameterization by Eq.~(7) in \cite{Fujimoto:2022ohj}
(see Eq.~(\ref{eq:LarryEoS}) below for three values of the parameter $\eta_c$ 
(read dashed/solid/dotted curves for $\eta_c = 1.2$, 1.1 and 1.0).
\label{fig:Larry} 
}
\end{figure}

An example of an EoS largely compatible with neutron star data is NY$\Delta$ 
based on the DD-ME2 density functional \cite{Li:2019fqe}.
Despite some peculiarities (see left panel in Fig.~\ref{fig:Larry}) 
it shares features recently advocated as essential w.r.t.\ QCD tracy anomaly and conformality
\cite{Fujimoto:2022ohj,Marczenko:2022jhl}.
A suitable measure of the trace anomaly is
\begin{align} \label{eq:Delta}
\Delta := \frac13 - \frac{p}{e}
\end{align}
which is related to the sound velocity squared
\begin{align} \label{eq:vs2}
v_s^2 := \frac{\partial p}{\partial e} = \frac{n_B}{\mu_b \chi_B} = \frac13 - \Delta - e \frac{\partial \Delta}{\partial e} ,
\end{align}
where $n_B$ stands for the baryon density, $\mu_B$ the baryo-chemical potential, and 
$\chi_B = \partial^2 p / \partial \mu_B = \partial n_B / \partial \mu_B$ is the second-order cumulant of the
net-baryon number density. Thermodynamic stability and causality constrain $\Delta \in [-2/3, 1/3]$.
Restoration of scale invariance means $\Delta \to 0$ and $v_s^2 \to 1/3$. Although $\Delta$ approaches monotonically to
zero, $v_s^2$ can develop a peak at lower energy densities. In fact, $v_s^2$ as a function of $\eta \equiv \ln e/(n_0 m_p)$
displays such peak at $\eta \approx 1.3$ (see Fig.~2 in \cite{Fujimoto:2022ohj}
or Fig.~1 in \cite{Altiparmak:2022bke} and discussion in \cite{Hippert:2021gfs}), 
which should be considered a signature of conformality even at strong coupling.
Continuing with the adjustment of $\Delta$ beyond the values tabulated in \cite{Li:2019fqe}
(see symbols in Fig.~\ref{fig:Larry}-left), $v_s^2$ drops first slightly below $1/3$ and approaches then slowly $1/3$
from below, in agreement with QCD asymptotic.

The comparison of the resulting mass-radius relations is exhibited in Fig.~\ref{fig:Larry}-right.
We follow the parameterization \cite{Fujimoto:2022ohj}
\begin{align} \label{eq:LarryEoS}
\Delta = \frac13 - \frac13 \left(1 - \frac{A}{B + \eta^2} \right) \frac{1}{1 + e^{- \kappa (\eta - \eta_c)}} ,
\end{align}
however, with parameters adjusted to \cite{Li:2019fqe}:
$\kappa = 3.45$, $\eta_c = 1.1$, $A = 2$ and $B = 20$,  see Fig.~\ref{fig:Larry}.
Despite the tiny differences of the red curve (for adjusted $\Delta$) and the solid black curve with markers
(for NY$\Delta$ DD-EM2 EoS of \cite{Li:2019fqe}), the overall shapes of $M(R)$ differ in detail,
see Fig.~\ref{fig:Larry}-right.
While the maximum masses agree, the radii for NY$\Delta$ DD-EM2 EoS of \cite{Li:2019fqe} are greater by
about 1~km at $M \approx 1.4~M_\odot$. Small dropping of the parameter $\eta_c$ from 1.2 to 1.0
in the $\Delta$ EoS let increase the radius by 1~km. All these differences can be traced back essentially  
to differences in the low-density part of the employed EoSs.

\subsection{Distinguished cores with NY$\Delta$ envelope}
 
\begin{figure}[t!]
\includegraphics[width=1.0\columnwidth]{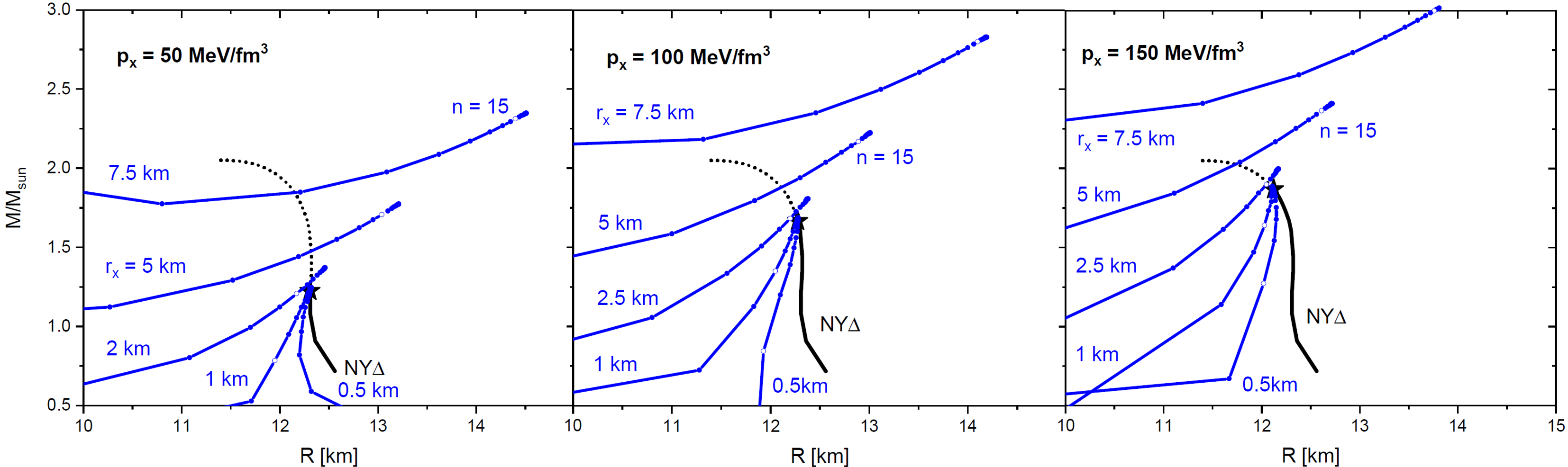}  %%  \hspace*{-1.45cm}
\caption{Mass-radius plane and its occupancy by compact stars with given 
core radii $r_x$ (blue curves with dots at core masses $m_x = 10^{-4} 1.5^n M_\odot$, 
$n = 1, 2, 3 \cdots$ from right to left; the \blau{open} circles depict points for $n = 15$\blau{, and the right-most endpoints are for $n=1$}). 
The values of $p_x$ are 50 (\blau{left} panel), 100 (middle panel) and 150~MeV/fm${}^3$ (\blau{right} panel). 
The fat solid curve is obtained by standard integration of the TOV equations
using the NY$\Delta$ EoS tabulated in \cite{Li:2019fqe}
with linear interpolation both in between the mesh and from the tabulated minimum energy density to the
$p=0$ point at $e_0 = 1$~MeV/fm${}^3$. The asterisks display the mass-radius values for 
$p_c = p_x$. That is, the sections above the asterisks (dotted curves)
are for a particular continuation of NY$\Delta$ at $p > p_x$,
which is, (trivially) in this case, NY$\Delta$ itself. One could instead employ 
the parameterization Eq.~(\ref{eq:coreEoS}) which would deliver another dotted curve. \blau{For other examples, in particular the small-$R$ region near black hole and Buchdahl limits, the interested reader is referred to \cite{Zollner:2022dst}.}
\label{fig:3} 
}
\end{figure}

Focusing now on NY$\Delta$ DD-EM2 EoS of \cite{Li:2019fqe} and the core-corona decomposition
one obtains the resulting mass-radius radius relation exhibited in Fig.~\ref{fig:3}. 
We find it convenient to keep the respective core radius constant and vary the core masses as
$m_x = 10^{-4} 1.5^n M_\odot$, $n = 1, 2, 3 \cdots$. The very small core masses\footnote{
Of course, a large core with several-km radius and small included mass is a very exotic thing;
it may be referred to as bubble or void with surface pressure $p_x$ and stable interface to SM matter.}
occupy the
right end of the solid blue curves, where the dots refer to increasing values of $n$. Heavy core masses
occupy the left sections of the blue curves, i.e.\ larger values of $n$.
For considerably smaller values of the core radii (not displayed), 
the blue curves approach the conventional mass-radius curve
of the NY$\Delta$ DD-EM2 EoS of \cite{Li:2019fqe} (fat black curve). The limit $r_x \to 0$ and $m_x \to 0$
of the core-corona decomposition curve is depicted by the asterisk, which also agrees with mass and radius
of NY$\Delta$ DD-EM2 EoS of \cite{Li:2019fqe} with $p_c = p_x$. Increasing values of $p_x$ make an up-shift 
of the core-corona mass-radius curves for a constant value of the core radius $r_x$, and a larger range of
masses is occupied. 

All the features discussed in Section~II in \cite{Zollner:2022dst} are recovered, e.g.\ the convex shape of the curves $r_x = const$, which however becomes visible only by extending the plot towards smaller values of $R$ (not shown).
The right end points of the core-corona curves refer to very small values of $m_x$,
while the not displayed left end points are on the limiting \blau{black hole} curve $2 G_N M = R$.
However, the core-corona decomposition shows that the maximum-mass
region of NY$\Delta$-DD-ME2 (see fat solid curves)
is easily uncovered too, interestingly with sizeable core radii and  
noticeably smaller up to larger total radii.

We refrain from displaying the core-corona mass-radius curves for $m_x = const$.
They can be easily inferred by connecting the points $n = const$ in each of the panels in Fig.~\ref{fig:3}.

To stress the relation to the usual mass-radius relation $M(R)$ obtained from the parametric representation
$M(p_c)$ and $R(p_c)$ let us mention that the \blau{respective dotted curve} sections above the asterisk on the fat solid curves
is just an example - here simply NY$\Delta$ DD-EM2 EoS of \cite{Li:2019fqe} for $p > p_x$.
Other continuations of the EoS at $p > p_x$ are within the region filled by the blue core-corona curves,
which however may not be completely mapped out by many conceivable EoSs: The core can contain more
than just SM matter, such as Dark Matter or Mirror Matter or other exotic material.
In the simplest case, a multi-component composition with hypothesized EoS for each component can be used
for the explicit construction. All what counts in our core-corona decomposition is that a core with radius $r_x$
and included mass $m_x$ is present and supports the pressure $p_x$ at its boundary. By definition,
outside the core, only SM matter with known EoS is there.
(For a dedicated study of crust properties, cf.\ \cite{Suleiman:2021hre}.)

\begin{figure}[t!]
\includegraphics[width=0.49\columnwidth]{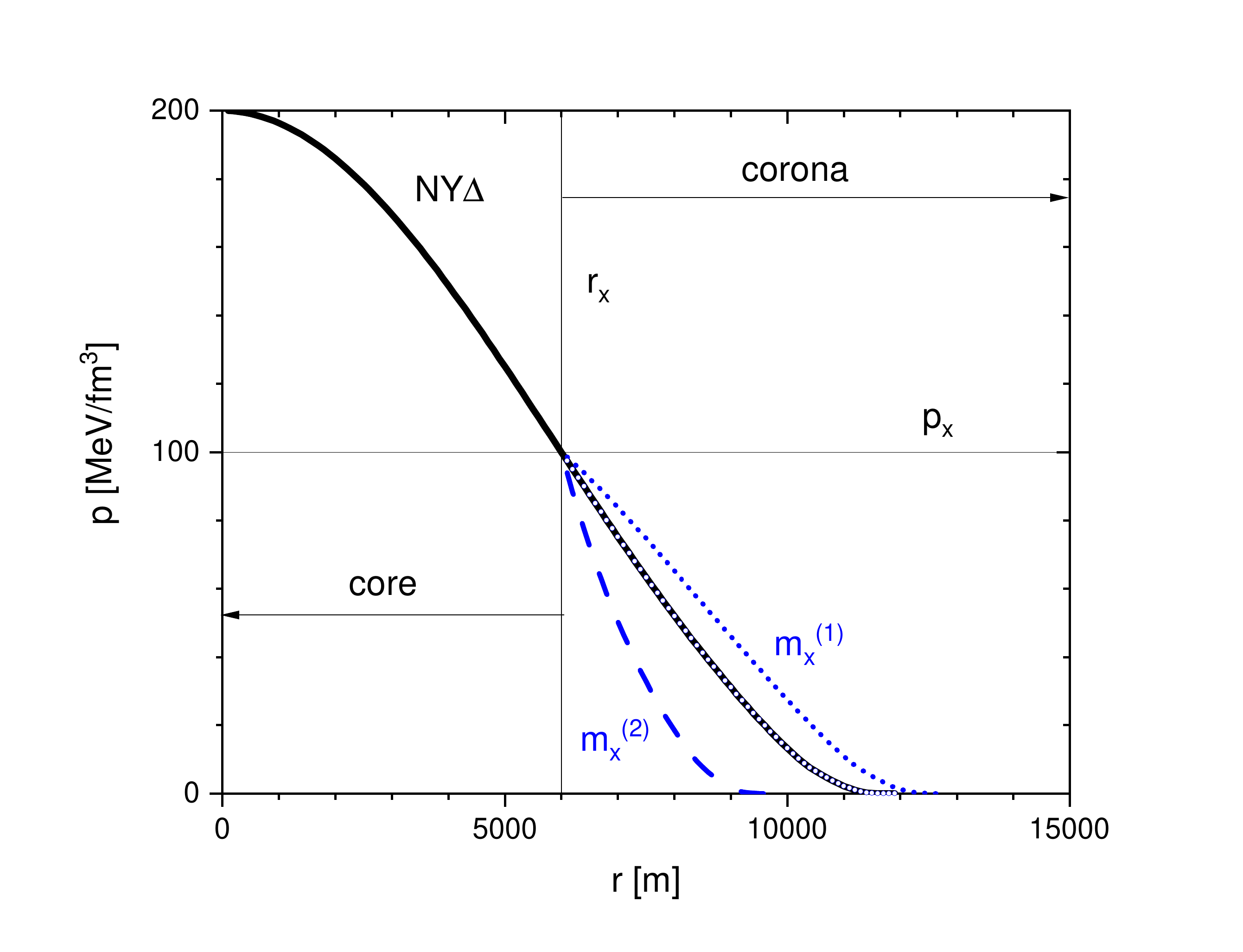}  % \hspace*{-9mm}
\includegraphics[width=0.49\columnwidth]{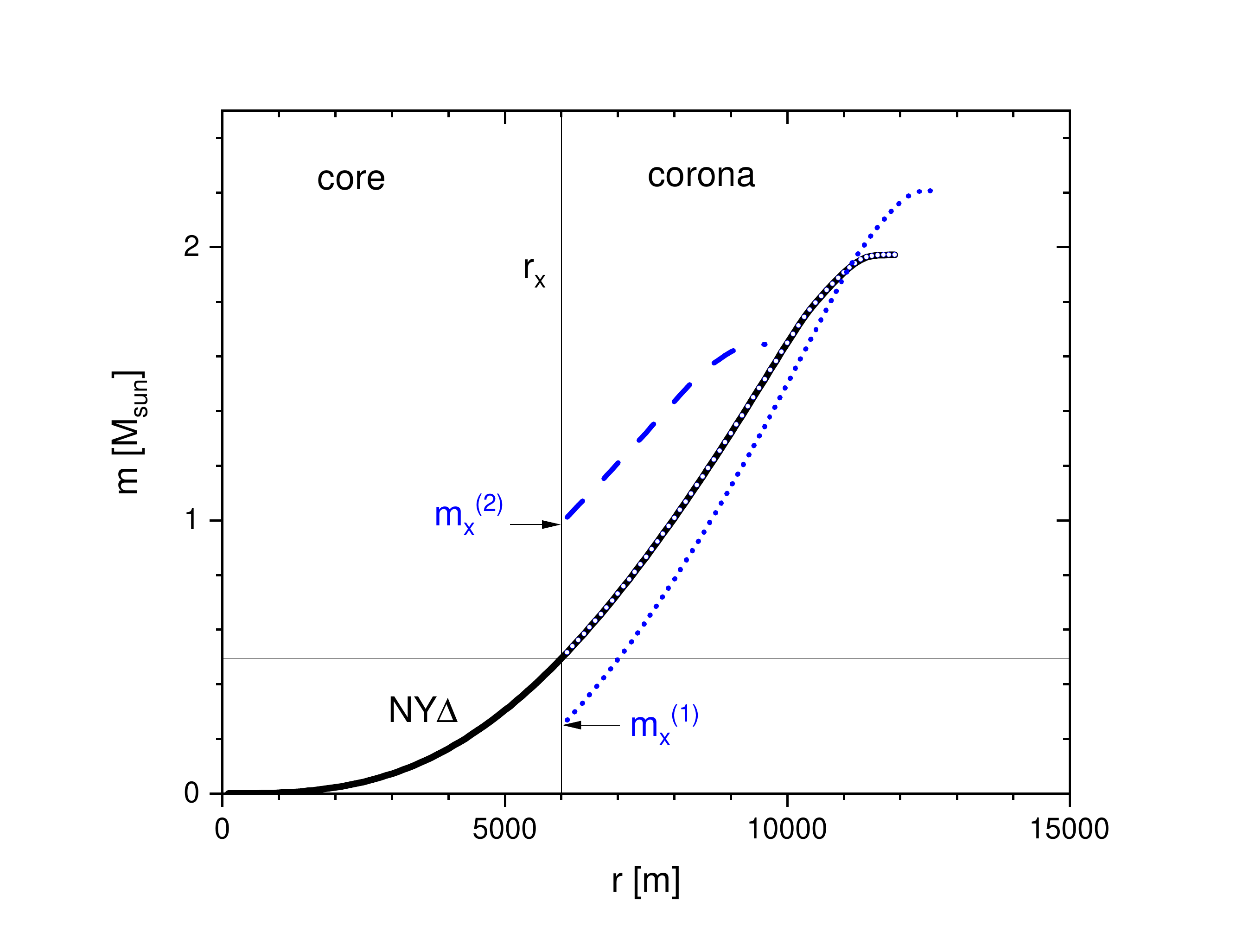}
\caption{Pressure $p(r)$ (\blau{left} panel) and mass $m(r)$ (\blau{right} panel) as a function of the radius $r$ for the special value
$p_x = 100$~MeV/fm${}^3$, selected here as end point of the ``reliable EoS'' NY$\Delta$
at $p \le p_x$. Assuming a possible continuation at $p > p_x$ by NY$\Delta$ itself as a particular example,
yields the fat solid curves for the ({\it ad hoc}) choice $p_c = 200$~MeV/fm${}^3$. Keeping the resulting
value $r_x$ from $p(r_x) = p_x$ and integrating the TOV equations in the corona,
$r \ge r_x$ with $p_x$ and $m_x^{(1,2)}$ as initial values, one gets the 
dashed blue ($m_x^{(1)} = 0.5 m_x$)
and dotted blue ($m_x^{(2)} = 2 m_x$) curves, 
where the respective values of $R$ and $M$ can be read off.
Using $m_x$ as core mass, the blue circles (on top of the fat black curve at $r \ge r_x$) are obtained.
Using a multitude of values $m_x^{(n)}$ would generates one additional blue curve in Fig.~\ref{fig:3}.
In the present example, for $r_x = 6$~km. 
\label{fig:4} 
}
\end{figure}

\subsection{Example of radial pressure and mass profiles}

To illustrate that feature in some detail let us consider an example and select $p_x =100$~MeV/fm${}^3$.
Assuming that the continuation of the EoS above this $p_x$ is, hypothetically, by NY$\Delta$ itself, one obtains 
the pressure and mass profiles as displayed \blau{by black solid curves} in Fig.~\ref{fig:4} for $p_c = 200$~~MeV/fm${}^3$.
Clearly, taking $r_x$, $m_x$ and $p_x$ as start values and integrating in the corona up to the surface at
$p=0$, one obtains the same values of $M$ and $R$ as for standard integration from $p(r=0) = p_c$ to
$p(R) = 0$ \blau{(see blue circles on top of the black solid curve sections)}. However, keeping $r_x$ but using other core masses $m_x$, e.g.\ $m_x^{(1)} = 0.5 m_x$
or $m_x^{(2)} = 2 m_x$, one obtains different pressure and mass profiles and, consequently, also
different values of $M$ and $R$, see Fig.~\ref{fig:4}. 
This is the very construction of the core-corona decomposition leading to the results exhibited in Fig.~\ref{fig:3}.

\section{Summary} \label{sect:summary}

The core-corona decomposition relies on the assumption that the EoS of compact/ neutron star matter 
(i) is reliably known up to energy density $e_x$ and pressure $p_x$ and 
(ii) occupies the star as the only component at radii $r \ge r_x$. 
The base line for static, spherically symmetric configurations is then provided by the TOV equations,
which are integrated, for $r \in [r_x, R]$, to find the circumferential radius $R$ (where $p(R) = 0$) and the
gravitational mass $M = m(R)$. We call that region $r \in [r_x, R]$ ``corona'', 
but ``crust'' or ``mantle'' or ``envelope'' or ``shell''
are also suitable synonyms. The region $r \in [0, r_x]$ is the ``core'', which is parameterized by 
the included mass $m_x$. The core must support the corona pressure at the interface, i.e.\ $p(r_x^-) = p(r_x^+)$.
The core can contain any material compatible with the symmetry requirements.
In particular, it could be modeled by a multi-component fluid with SM matter plus Dark Matter and/or Mirror World matter
or anything beyond the SM. The TOV equations deliver then $R(r_x, m_x, p_x)$ and $M(r_x, m_x, p_x)$.

It helps our intuition to think of a SM matter material in the core with assumed fiducial EoS which determines, for given
value of $p_x$, $r_x(p_c)$ and $m_x(p_c)$, thus $R(p_c)$ and $M(p_c)$ resulting in $M(R)$, as conventionally done.  
Using many fiducial test EoSs at $p > p_x$, one maps out a certain region in the $M$-$R$ plane
accessible by SM matter. Our core-corona decomposition extends this region, since the core can contain
much more than SM matter only. With reference to a special EoS, applied tentatively in core and corona, the accessible
masses and radii become smaller (larger) for heavy (light) cores. That effect is noticeable for large (km size)
and heavy (fraction of solar mass) cores. Small and light cores have hardly an impact. 

This analysis should be
refined in follow-up work by employing improved EoSs in the low-density region and by taking care of 
mass-radius values for heaviest neutron star(s) still keeping precise radius values for the $1.4 M_\odot$ neutron stars.  
Also the holographic model in Appendix~\ref{sect:holography}, which \blau{is aimed at mapping} hot QCD thermodynamics to a cool EoS, deserves further investigations\blau{, before arriving quantitatively at an EoS suitable for compact star properties and (merging) dynamics. The emphasis in Appendix~\ref{sect:holography} is the illustration of the extrapolation scheme beyond a patch of QCD thermodynamic state space.}  

Taking up the suggestion in \cite{Baym:2017whm} that the scaling property of the TOV equations
with a mass dimension-4 quantity, e.g.\ by $m_p^4$, where $m_p$ is the proton mass,
determines essentially gross parameters of compact/neutron stars, we supply \blau{in Appendix~\ref{sect:A}} a brief contemplation on the emergence
of hadron masses within QCD-related approaches, thus bridging from micro to macro physics:
$M_{PSR \, J0740+6620} \approx 2 M_\odot \approx 2.38 \times 10^{57} m_p$,
$R_{PSR \, J0740+6620} \approx 12.3~\mbox{km} \approx 1.5 \times 10^{19} r_p$
with proton charge radius $r_p \approx 0.88~\mbox{fm} \approx 4.2 m_p^{-1}$
which is somewhat larger than its ``natural'' scale $\hbar c  / m_p$.
Our focus, however, is here on the emergence of the gravitational mass\footnote{
The gravitational mass in our core-corona decomposition is
$M = m_x + 4 \pi \int_{r_x}^R \dd r r^2 e(r)$ which, due to gravitational binding, is different from the total mass 
$\int \dd^3 V e(r)$ where
the integral measure d$^3 V$ accounts for the gravitational spatial deformation, cf.\ \cite{Gao:2019vby}.}  
determining the trajectories of light rays and test particles outside of compact astrophysical objects. 

%%%%%%%%%%%%%%%%%%%%%%%%%%%%%%%%%%%%%%%%%%

%%%%%%%%%%%%%%%%%%%%%%%%%%%%%%%%%%%%%%%%%%
%\authorcontributions{Conceptualization, B.K.; methodology, R.Z., M.D. and B.K.; validation, R.Z., M.D. and B.K.; investigation, R.Z., M.D. and B.K.; writing---original draft preparation, B.K.; writing---review and editing, B.K. and R.Z.; visualization, B.K. and R.Z.; supervision, B.K.; project administration, B.K. All authors have read and agreed to the published version of the manuscript.}

%\funding{The work is supported in part by the European Union’s Horizon 2020 research and innovation program STRONG-2020 under grant agreement No 824093.}

\acknowledgments{We thank S.~M.~Schmidt and C.~D.~Roberts for inviting our contribution to the topic of the emergent mass phenomenon. M.~Ding is grateful for support by Helmholtz-Zentrum Dresden-Rossendorf High Potential Programme. One of the authors (BK) acknowledges continuous discussions with J.~Schaffner-Bielich and K.~Redlich and for the encouragement to deal with the current topic. The work is supported in part by the European Union’s Horizon 2020 research and innovation program STRONG-2020 under grant agreement No 824093.}

%\conflictsofinterest{The authors declare no conflict of interest.} 

%%%%%%%%%%%%%%%%%%%%%%%%%%%%%%%%%%%%%%%%%%
%% Optional
%\appendixtitles{yes} % Leave argument "no" if all appendix headings stay EMPTY (then no dot is printed after "Appendix A"). If the appendix sections contain a heading then change the argument to "yes".
%\appendixstart
\appendix
\section{Emergence of hadron masses}\label{sect:A}

Astronomical observations of neutron stars have provided us with a way to understand these mysterious objects. The literature suggests that the outer and inner core of neutron stars may consist of nucleons or cold, dense quark matter. The question one can ask is how almost massless light quarks could combine and produce such massive nucleons to give rise to neutron star \blau{mass, as emphasized in Subsection \ref{sect:2-1}.} A widely recognized mechanism is the Higgs mechanism, which gives rise to the current quark masses. However, as was immediately realized, the current quark mass contributes only a few MeV, while the mass of a proton or neutron is much larger. The mystery of the missing masses lies in the emergence of hadron mass (EHM).

\subsection{Three pillars of EHM}

Contemporary studies of continuum Schwinger methods (CSMs) have shown that the emergence of hadron mass \blau{rests} on three pillars: (a) the running quark mass, (b) the running gluon mass, and (c) the process-independent effective charge. These three pillars provide the basis for giving observable results \blau{of} hadron properties~\cite{Ding:2022ows}. 

\subsubsection{Running quark mass}

The first pillar is the most familiar, namely the running quark mass. The dressed-quark propagator can be represented by a special quantity, the dressed-quark mass function. Modern CSMs calculations of the quark gap equation show that the dressed-quark mass function is a finite value of $M_0(k^2=0)=0.41$ GeV in the far infrared, even in the chiral limit. This is an explicit expression of the dynamical chiral symmetry breaking (DCSB), and the infrared scale is responsible for the masses of all hadrons, and the running quark mass is therefore regarded as an expression of the EHM. In a sense, the quark in the far infrared can be seen as a quasiparticle, produced by the interaction of the high-energy quark parton with the gluon. The scale of the dressed-quark mass function in the far infrared is comparable to the scale in the constituent quark model. The difference is that the dressed-quark mass function is momentum-dependent, and it runs from the far infrared to the ultraviolet, where it matches the current quark mass in perturbation theory. The dressed-quark mass function is also flavor-dependent, with the Higgs mechanism gradually dominating in describing it from light to heavy quarks, while the dressed-quark mass function decreases at a slower rate. It is worth emphasizing that the ultraviolet behavior of the dressed-quark mass function in the chiral limit is related to the well known chiral condensates.

\subsubsection{Running gluon mass}

The second pillar is the running gluon mass. Gauge invariance requires that the gluon parton be massless. However, in the Schwinger mechanism of QCD, it was conjectured long ago that the two-point gluon Schwinger function might give rise to a massive dressed gluon. Thus, the dressed gluon gains mass and the gluon parton is transformed into the gluon quasiparticle by interaction with themselves. Consequently, gluons can have a momentum-dependent mass function, and this non-zero function has been confirmed by both the continuum and lattice QCD. The dressed gluon mass function is power-law suppressed in the ultraviolet, so it is invisible in perturbation theory. However, it is non-zero in the infrared momentum \blau{range}, and in the far infrared it yields a value \blau{of} $m_g = 0.43$ GeV. This is purely a manifestation of ``mass from nothing'' and is responsible for all hadron masses. It is worth emphasizing that the running gluon mass may also be associated with confinement. The gluon two-point Schwinger function has an inflection point, so it has no K\"all\'en-Lehmann representation, and therefore the relevant states cannot appear in the Hilbert space of physical states. 

\subsubsection{Process-independent effective charge}

The third pillar is the process-independent effective charge. QCD has a running coupling which expresses the feature of asymptotic freedom in the ultraviolet at one-loop order, but it also shows a Landau pole in the infrared, where the coupling becomes divergent when $k^2=\Lambda_{\text{QCD}}^2$. However, recent advances in CSMs based on pinch techniques and background field methods show that QCD has a unique, nonperturbatively well-defined, computable, process-independent effective charge. Its large-momentum behavior is smoothly connected to the QCD running coupling at one-loop order, while it is convergent at small momentum. The Landau pole is eliminated due to the appearance of the gluon mass scale in the infrared. It is noteworthy that\blau{,} at small momentum\blau{,} the effective charge run ceases and it enters a domain that can be regarded as an effective conformal. In this domain, the gluons are screened so that the valence quasiparticle can be seen as carrying all the properties of a hadron. Furthermore, the process-independent effective charge matches the Bjorken process-dependent charge, and in practical use they can be considered to be indistinguishable. Additionally, the process-independent effective charge is a well-defined smoothing function \blau{at} all available moment\blau{a}, from the far infrared to the ultraviolet, so it can be a good candidate for \blau{applications}, for example, in evolving parton distribution functions to different scales.   

\subsection{Hadrons in vacuum}

A direct correlation with the running quark mass  is its effect on pseudoscalar mesons, especially the lightest meson, \blau{the} pion~\cite{Dorkin:2013rsa,Dorkin:2014lxa}. As a chiral symmetry breaking Nambu-Goldstone boson, the well-known 
Goldberger-Treiman relation relates the dressed-quark mass function to pion's Bethe-Salpeter amplitude, so that, to some extent, it can be seen that\blau{,} once the one-body problem is solved, the two-body problem is also solved. Furthermore, because of its direct connection to the dressed quark mass function, the properties of \blau{the} pion can be seen as the cleanest window to glimpse the emergence of hadron mass. In the quark model, the vector meson is a spin-flip state of the pseudoscalar meson, and thus the $\rho$ meson is the closest relative to pion\blau{s}. However, the properties of $\rho$ mesons are significantly different from those of pion. Compared to the surprisingly light pion, the $\rho$ is much heavier. Therefore, one can ask a straightforward question: how does the spin flip produce such different masses for the two mesons? \blau{Since} QCD is a well-defined theory of strong interactions, it must answer such a fundamental question. 

The meson spectrum includes not only ground states, but also excited states with high orbital angular moment\blau{a} between the quark and antiquark \blau{constituents} of the meson. The study of excited mesons is more difficult in both lattice QCD and CSMs. Experience tells us that the well-known rainbow ladder approximation does not provide the correct ordering of the meson masses between ground and excited states when using CSMs~\cite{Greifenhagen:2018ogb}. Recent improvements to the Bethe-Salpeter kernel have made it possible for practitioners of CSMs to reproduce the empirical results. The new feature of modern Bethe-Salpeter kernel is the inclusion of a term closely related to the anomalous chromomagnetic moment, which reflects the influence of the EHM \blau{on} the quark gluon vertex and which can be used to effectively describe the excited meson state very well.

In addition to studying mesons made up of light quarks, it is also worth exploring how the properties of mesons vary with meson mass. In the quark model view, if a valence light quark in pion is replaced by a strange quark, a kaon is formed. A kaon is also a Nambu-Goldstone boson in the chiral limit, however, in the real world, the strange quark\blau{,} produced through the Higgs mechanism\blau{,} is $27$ times more massive than the current light quarks. Thus, the properties of the kaon are the result of the combined effect of the Higgs mechanism and the emergence of hadron mass. This has been revealed from the study of the parton distribution functions of kaons. The skewness of the distribution is caused by the heavier strange quark, while the overall broadening of the distribution is caused by the EHM. If mesons consisted of heavier quarks, charm and bottom quarks, the Higgs mechanism would dominate and be the largest source of heavy meson properties. In particular, pseudoscalar mesons and \blau{heavy} vector mesons are of particular interest as mesons with zero orbital angular momentum in the quark model ~\cite{Dorkin:2010ut,Dorkin:2010pb}. Their distributions are usually narrower than those of light mesons, since it has been pointed out that the distribution in the heavy\blau{-}quark limit is a Dirac delta function. The difference in distributions provides a clear picture of how the properties of mesons evolve with increasing meson mass, and the CSMs is a unified framework for describing all mesons, from pion to Upsilon.

It is worth mentioning that progress in mesons has also been extended to baryons. The properties of baryons are calculated using the Faddeev equation describing the three-quark scattering problem. Since the complete three-body problem in nature is much more complicated, the quark dynamical-diquark method is usually introduced, which is useful as a means of elucidating many qualitative features. Central to this approach is the incorporation of five different diquark correlations, of which the axial-vector diquark is of outstanding importance, to produce the correct baryon spectrum as well as baryon structure functions such as distribution functions, electromagnetic, axial and pseudoscalar form factors.

In addition to the existing traditional hadrons, mesons and baryons, many other new hadron states have been proposed experimentally and theoretically, such as exotic states, pentaquark, tetraquark, hybrid states and glueballs. In the field of research on mesons consisting of heavy quarks, a comprehensive study of the charm family has been carried out thanks to a large amount of experimental data from the B-factory. As a result, states such as XYZ states, pentaquark and tetraquark have extensively extended our knowledge of QCD bound states. In the field of research on mesons consisting of light quarks, there is a tendency to think that gluon degrees of freedom may also play a role in the formation of bound states, and thus there is speculation about the existence of states such as hybrid states and even glueballs~\cite{Kaptari:2020qlt}. This has been found from calculations of lattice QCD and CSMs, and future experiments, such as the 12 GeV upgrade experiment at Jefferson Lab, will provide an opportunity to test these theoretical predictions.

\subsection{Hadrons in cold dense medium}

In \blau{a} cold and dense medium, i.e.\ for non-zero baryo-chemical potentials, 
quarks and emergent hadrons \blau{suffer the impact of the ambient matter.} From low to high chemical potential domains, QCD will go through a phase of confinement and dynamical chiral symmetry breaking to a phase of deconfinement and chiral symmetry restoration. The key to the study is to determine the critical point/region at which the transition occurs 
and to which category the ``phase transition'' belongs.

Since the full domain of the finite chemical potential is currently \blau{not} fully available in lattice QCD simulations, a complementary approach is CSMs~\cite{Maris:1997eg,Bender:1997jf,Roberts:2000aa}, which expresses the dynamical chiral symmetry breaking and confinement in QCD, so that it can therefore be used as a tool to explore the properties of quark and hadron matter in cold dense medium, thus revealing relevant features of objects such as neutron stars. In the CSMs, the medium-induced dressed-quark propagator can be obtained by solving the quark gap equation in the medium, and the quark condensates are proportional to the matrix trace of the dressed quark propagator in the chiral limit. The quark condensate is crucial because it is commonly seen as the order parameter of the \blau{deconfinement (phase)} transition. In some studies it has been shown that the \blau{chiral} quark condensate is discontinuous with chemical potential, so that in the chiral limit the phase transition is of first order. Furthermore, it has been suggested that \blau{the} QCD quark condensate may be completely contained in hadrons \cite{Brodsky:2012ku}. In addition to quarks, hadrons are also affected by the \blau{ambient} medium at a non-zero chemical potential and\blau{, consequently,} their masses change, showing \blau{noticeable} deviations from the masses in vacuum. The appearance of a turning point in the chemical potential dependence of the hadron mass can also characterize the occurrence of chiral symmetry transition. It has also been proposed that increasing the chemical potential promotes the possibility of quark-quark Cooper pairing, i.e. diquark condensation. Quark-quark Cooper pairs are composite bosons with both electric and color charge, and thus superfluidity in quark matter entails superconductivity and color superconductivity.

These hadron\blau{ic} in-medium information is important not only for imaging the QCD phase diagram, but also for constructing a unified EoS from low-density nucleonic matter to high-density quark matter, and thus becomes crucial for determining the properties of neutron stars. \blau{Analog reasoning applies to other forms of strong-interaction matter which could govern new classes of compact stars,
e.g.\ pion stars \cite{Brandt:2018bwq}.}

\subsection{Supplementary remarks}

Before supplementing the above reflections by another approach to EHM w.r.t.\ condensates 
as fundamental QCD quantities,
let us mention that the {\sl ab initio} access to hadron \blau{vacuum} masses is directly based 
on QCD as the theoretical basis of strong interaction. 
The current status is reviewed in \cite{Brambilla:2014jmp}: a compelling description of the
mass spectrum and various other hadron parameters is achieved by lattice QCD.

The operator product expansion relates parameters of a hadron model
of the spectral function to a series of QCD condensates,
most notably the \blau{above mentioned} chiral quark condensate $\langle \bar q q \rangle$,
the gluon condensate $\langle \frac{\alpha_s}{\pi} G^2 \rangle$,
the mixed quark-gluon condensate $\langle \bar q g_s \sigma  G q \rangle$,
the triple gluon condensate $\langle g_s^3 G^3 \rangle$,
the four-quark condensates 
$\langle  \bar q \Gamma q  \bar q \Gamma q \rangle$\footnote{
$\Gamma$  denotes all possible structures formed by Dirac and Gell-Mann matrices.} \cite{Thomas:2005dc}
and the poorly known condensates with higher mass-dimension.
Further ingredients are the Wilson coefficients which are accessible by perturbation theory \cite{Buchheim:2014rpa}.
In a strong-interaction medium, the condensates change:
$\langle \cdots \rangle_0 \to  \langle \cdots \rangle_{T, \mu_B}$ -- 
one could say that they are expelled from vacuum (label ``$0$'')
by a higher spatial occupancy due to non-zero temperature and density (parametrized by $\mu_B$).\footnote{
Similar to concerns as on the Higgs field condensate challenge such a picture of the vacuum, cf.\ \cite{Brodsky:2009zd}
and follow-up citations.}
The induced dropping of condensates facilitates in-medium modifications
of hadrons, in particular due the coupling to the chiral condensate, which is often considered as order
parameter of chiral symmetry.   
A recent review is \cite{Gubler:2018ctz}. 

In the baryon sector, the impact of the four-quark condensates and other in-medium-only condensates
make unambiguous analyses of QCD sum rules somewhat vague. Previous factorizations relate the many 
four-quark condensates to the chiral condensate modulo some uncertain factor. For a comprehensive
discussion, see \cite{Thomas:2007gx} which focuses on the emergence of the nucleon mass,
most notably $m_p$, from various quark and gluon condensates.

\begin{figure}
\includegraphics[width=0.99\columnwidth]{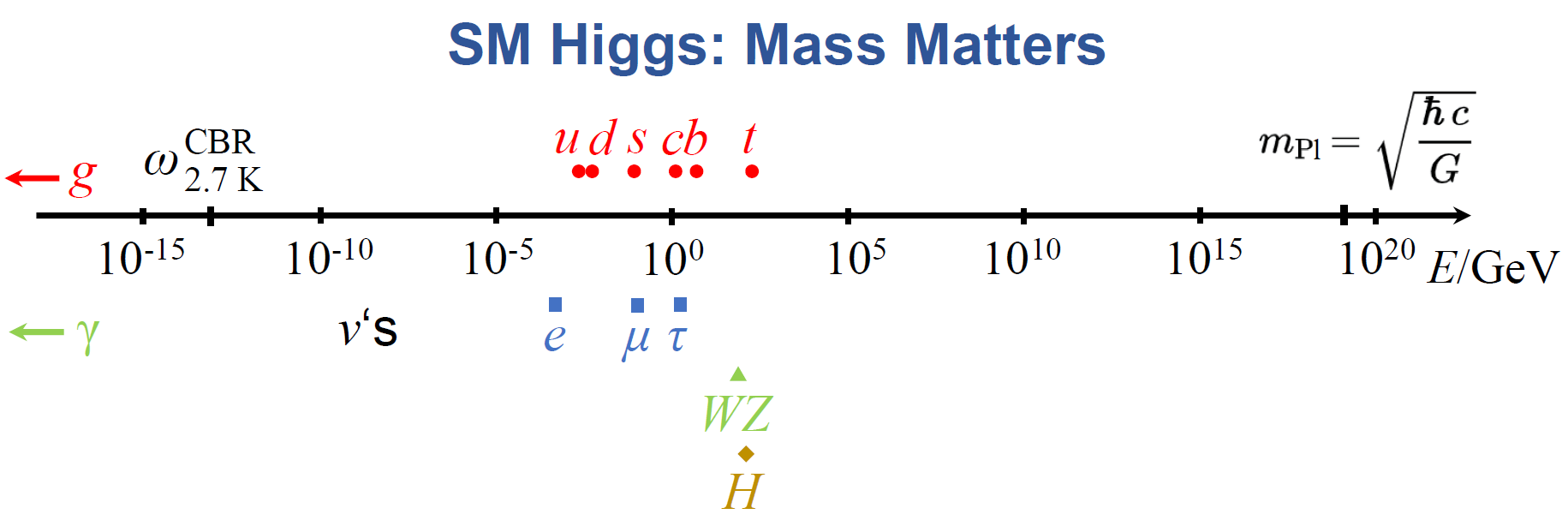} \\[-3mm]
\caption{Masses within the SM. Mysterious concentration of bare SM masses 
(quarks $[u, d, s, c, t, b]$, leptons $[e , \mu, \tau]$,
gauge bosons $[W^\pm, Z^0]$, Higgs $[H]$) and  
separation of neutrinos  ($\nu's = [\nu_{e , \mu, \tau}])$ 
on a large energy scale ranging from present-day cosmic background radiation 
$\omega_{2.7~K}^{CBR} \approx 0.233 \times 10^{-3}$~eV
to Planck mass $m_{Pl} = \sqrt{\hbar c/G_N} \approx 1.22 \times 10^{19}$~GeV.
Only the QED and QCD gauge Bosons $[\gamma, g]$ remain massless.  
\label{fig:Higgs} 
}
\end{figure}

This brief journey should supplement our perspective of how the hadronic constituents of matter in compact stars acquire
their masses and that the ambient medium modifies them. The very first step of gaining the above mentioned
bare input masses by the Higgs mechanism is thereby left out, see Fig.~\ref{fig:Higgs}.
The various bare quark masses appear in several QCD approaches as parameters adjusted to data.
This is evidenced, e.g.\ in the approach \cite{Greifenhagen:2018ogb}:
A formula of two-quark meson masses is developed,
$m(m_1, m_2)$, where $m_{1, 2}$ refer to the bare quark masses.
% entering the Dyson-Schwinger equation with a unique interaction kernel. 
As shown in Fig.~3 in \cite{Greifenhagen:2018ogb}, inspection of the 
contour lines $m(m_1, m_2) = const.$ can be used to pin down numerical values of bare masses $m_{1, 2}$
which reproduce empirical values of ground states in the pseudo-scalar and vector channels including light
$u, d$, strange and charm quarks at once; redundancy is used for cross checking.
Also Regge trajectories become accessible to some extent \cite{Greifenhagen:2018ogb,Fischer:2014xha}.

\section{Holographic approach to the EoS}\label{sect:holography}

Here we present a particular model of strong-interaction matter which is based on the famous AdS/CFT correspondence.
In line with \cite{DeWolfe:2010he,Grefa:2021qvt,Critelli:2017oub} 
we employ the action in a fiducial five-dimensional space-time with asymptotic AdS symmetry:
\begin{align}
S_{EdM} &= \frac{1}{2 \kappa_5^2} \int \dd^4 x \, \dd z \, \sqrt{-g_5}
\left( R - \frac12 \partial^M \phi \, \partial_M \phi - V(\phi) 
 - \frac14 {\cal G} (\phi) F_{{\cal B}}^2 \right) , % + S_{GH},
\label{EdM_action}
\end{align}
where $R$ is the Einstein-Hilbert gravity part,
$F_{\cal B}^{MN} = \partial^M {\cal B}^N - \partial^N {\cal B}^M$
stands for the field strength tensor of an Abelian gauge field ${\cal B}$ \`{a} la Maxwell
with ${\cal B}_M \dd x^M = \Phi (z) \, \dd t$  defining the electro-static potential.
An embedded black hole facilitates the description of a hot and dense medium, since the black hole has
a Hawking surface temperature and sources an electric field, thus encoding holographically 
a temperature and a density of the system.
Dynamical objects are a dilatonic (scalar) field $\phi$ and a Maxwell-type field $\Phi$
which are governed by a dilaton potential
$V(\phi)$ and a dynamical coupling ${\cal G}(\phi)$ and geometry-related quantities.\footnote{
For a more general holographic approach to compact star physics, cf.\ \cite{Jarvinen:2021jbd}.}
Space-time is required to be described by the line element squared
\begin{align} \label{eq:3}
\dd s^2 = g_{MN} \dd x^M \dd x^N :=
 \exp\{ A(z, z_H)\} 
\left[f(z, z_H) \, \dd t^2 - \dd \vec x^2 - \frac{\dd z^2}{f(z, z_H)} \right] ,
\end{align}
where $z_H$ is the horizon position, $z \in [z_H, \infty]$ the radial coordinate, $A$ the warp factor and $f$ the blackness function.

\begin{figure*}[t!]
\includegraphics[width=0.31\columnwidth]{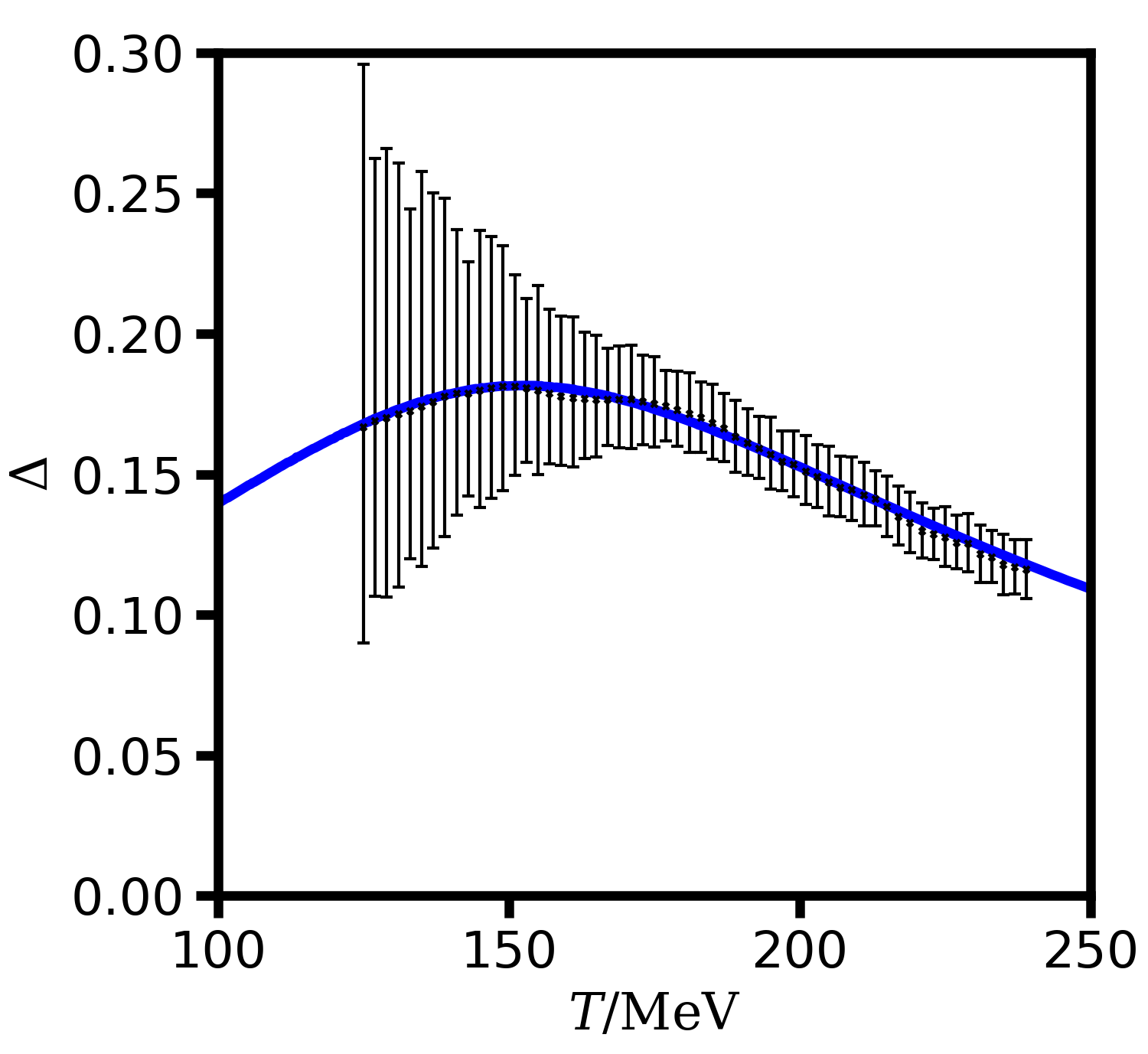}
\includegraphics[width=0.31\columnwidth]{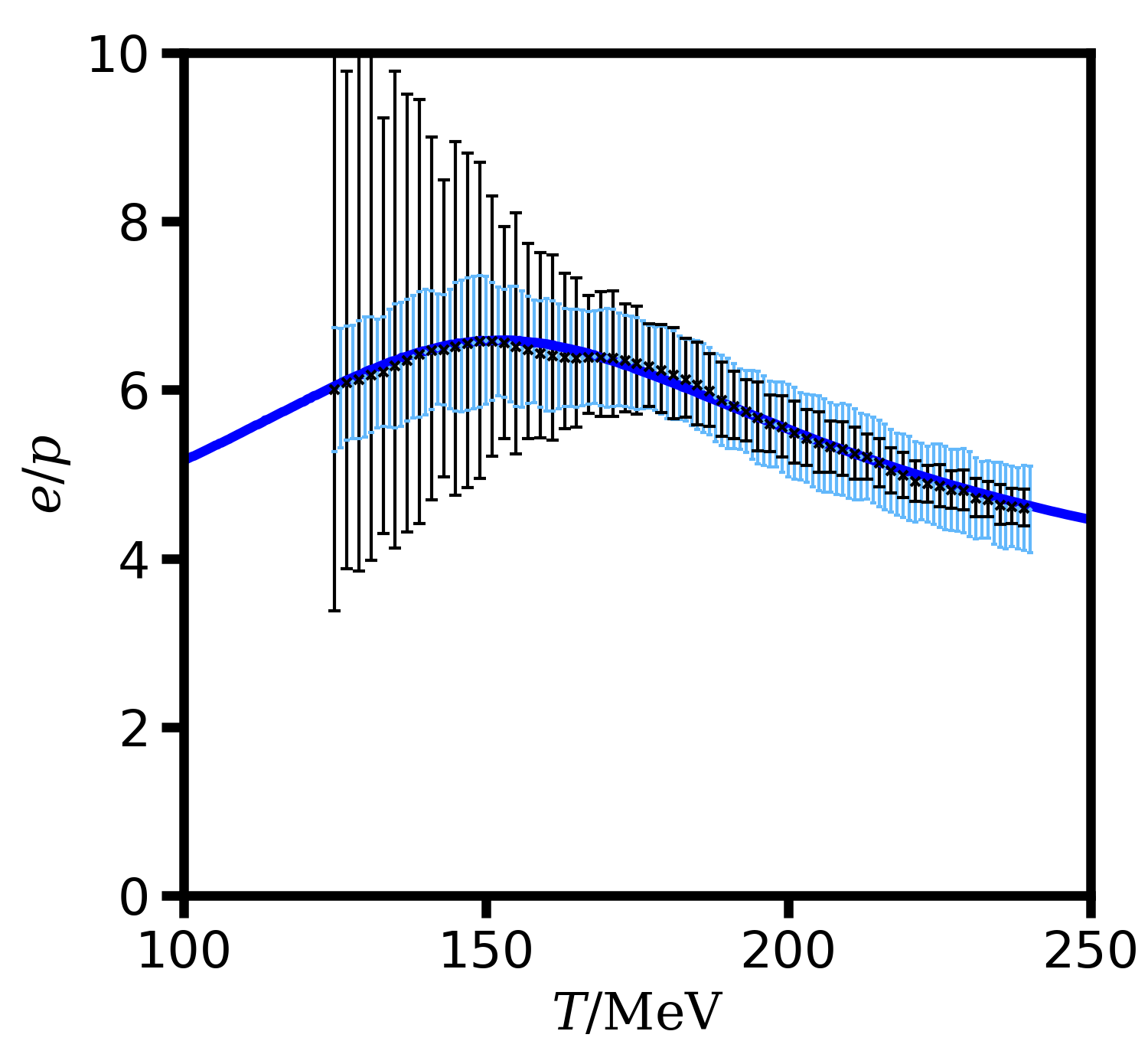}
\includegraphics[width=0.31\columnwidth]{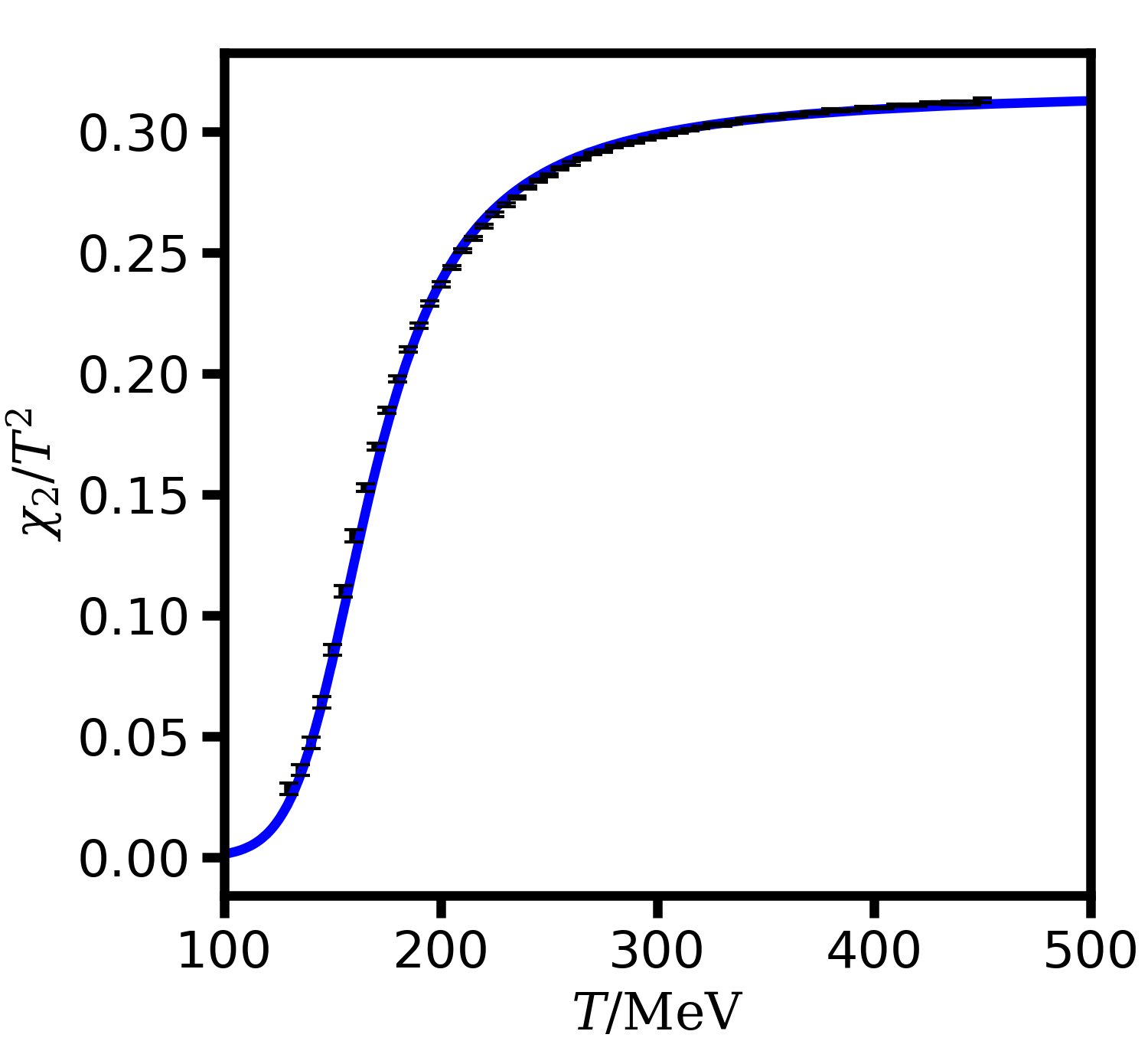}
\caption{Trace anomaly measure $\Delta$ (left panel) and ratio $e/p$ (middle panel)
for the holographic model with tuned parameters to describe the lattice QCD data \cite{Borsanyi:2021sxv} 
(small crosses) at $\mu_B = 0$.
Errors are constructed either from combining respective maximum and minimum values (vertical error bars)
or by error propagation in quadrature (blueish error bars). 
The scaled susceptibility $\chi_2 /T^2$ is displayed in the right panel; data (symbols) from \cite{Bellwied:2015lba}.
\label{fig:Delta} 
}
\end{figure*}

\begin{figure*}[t!]
\includegraphics[width=0.31\columnwidth]{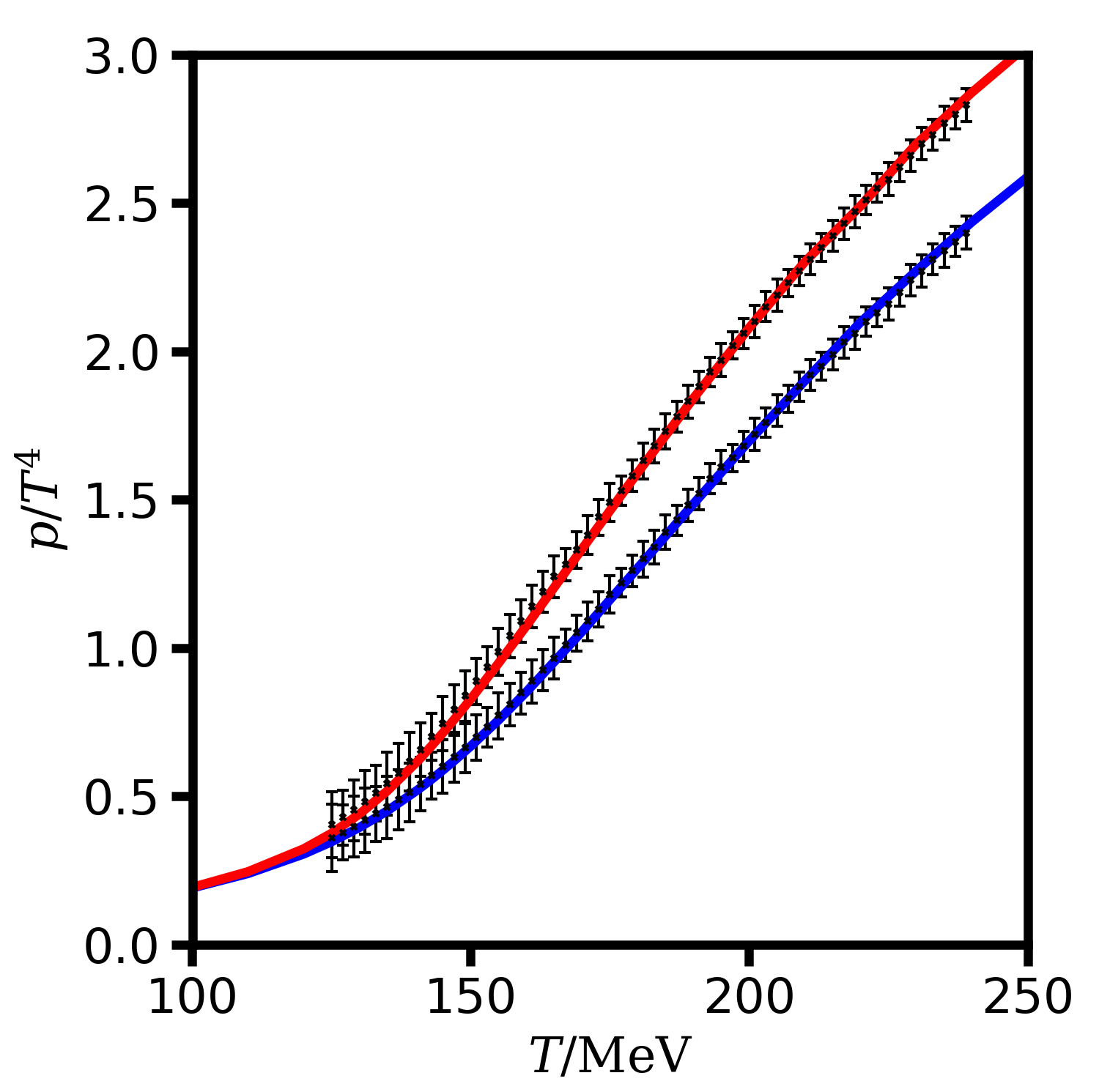}
\includegraphics[width=0.31\columnwidth]{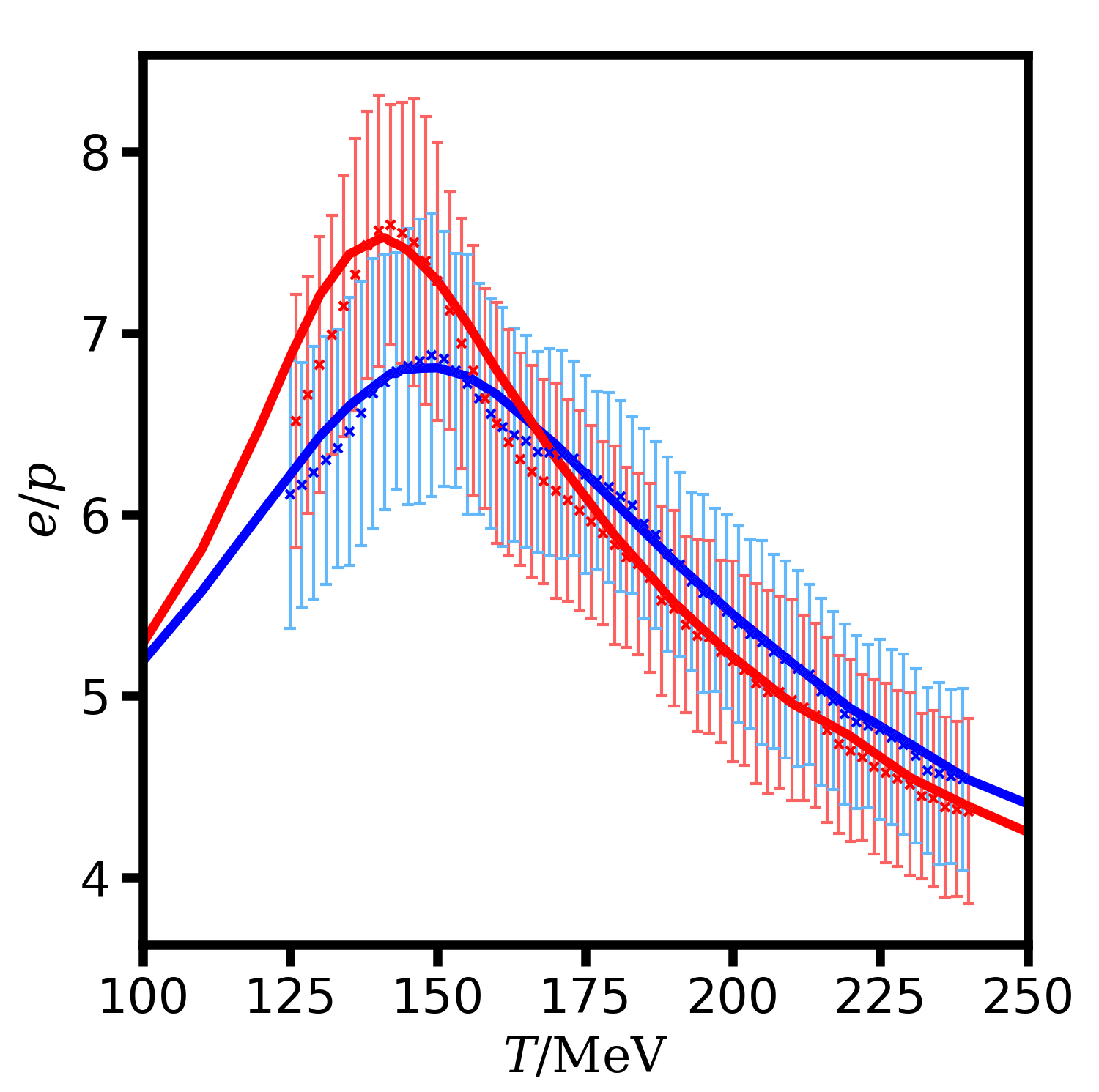}
\includegraphics[width=0.31\columnwidth]{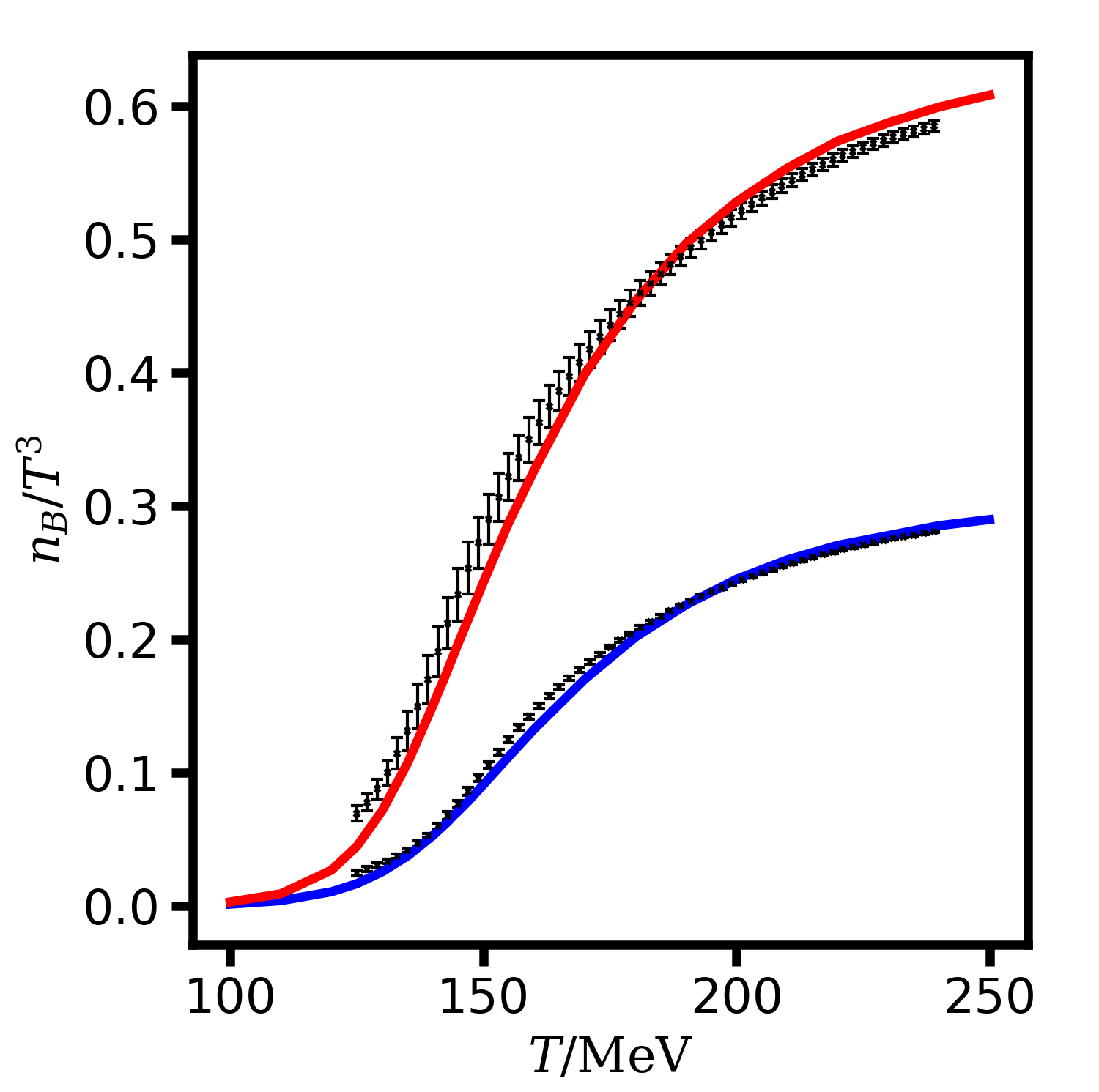}
\caption{Scaled pressure $p/T^4$ (left panel), ratio $e/p$ (middle panel) and scaled baryon density $n_B/T^3$ (right panel) as a function of temperature for $\mu_B /T = 1$ (blue) and 2 (red) in comparison with the data \cite{Borsanyi:2021sxv} (symbols with error bars).
\label{fig:new_hol_plots} 
}
\end{figure*}

The resulting Einstein equations are a set of coupled second-order ODEs to be solved with appropriate boundary conditions.
The quantities $V$ and ${\cal G}$ are tuned\footnote{
In contrast to former work, we here put emphasis on side conditions which ensure that, at $\mu_B = 0$,
no phase transition is facilitated outside the temperature range uncovered by the lattice data.}
to describe quantitatively the lattice QCD results 
\cite{Borsanyi:2021sxv}:\footnote{
While, at $\mu_B = 0$, the data sets \cite{Borsanyi:2013bia,HotQCD:2014kol} are consistent, special combinations
of quantities, e.g.\ $e/p$, enhance small differences.
This is, in particular, striking at $T \in [130, 140]$~MeV, where a $+4$\% ($-10$\%) deviation 
in energy density (pressure) even changes the shape of the $e/p$ curve when ignoring error bars.}
\begin{align}
\partial_\phi \ln V(\phi) &= (p_1 \phi + p_2 \phi^2 + p_3 \phi^3 ) \exp\{ - \gamma \phi\} , \label{eq:dil_pot}\\
{\cal G} (\phi) &= \frac{1}{1 + c_5 } \left( \frac{1}{\cosh (c_1 \phi + c_2 \phi^2)}
+ \frac{c_3}{\cosh c_4 \phi} \right) \label{eq:dyn_coup}
\end{align}  
with parameters $\{ p_{1,2,3}, \gamma \} = \{ 0.165919, $ $0.269459, $ $-0.017133, $ $0.471384 \}$
and $\{ c_{1, 2, 3, 4, 5}\} = \{-0.276851, $ $0.394100, $ $0.651725, $ $101.6378, $ $-0.939473 \}$.
These parameters and the scales refer implicitly to the QCD input, see Fig.~\ref{fig:Delta}.
The trace anomaly measure $\Delta$, Eq.~(\ref{eq:Delta}), is exhibited in Fig.~\ref{fig:Delta}-left for various temperatures $T$
and at baryon-chemical potential $\mu_B  = 0$; the middle panel is for the ratio $e/p$.
As argued in \cite{Fujimoto:2022ohj}, the high-temperature, small-density
and low-temperature, high-density behavior is strikingly different.
A further crucial input quantity is the susceptibility $\chi_2 = \partial_{\mu_B} n_B \vert_{\mu_B = 0}$,
see right panel in Fig.~\ref{fig:Delta}.
The length scale of the $z$ coordinate is set by $L^{-1} = 1465$~MeV,
which relates the horizon position $z_H$ and temperature via 
$T = - 1 /(4 \pi \partial_z f (z, z_H) )\vert_{z = z_H}$,
and by $\kappa_5^2 = 8.841$~fm${}^{-3}$, which determines
entropy density $s(T, \mu_B) = \frac{2 \pi }{\kappa_5^2} \exp\{\frac32 A(z_H, z_H)\}$
and baryon density $n_B (T, \mu_B) = - \frac{\pi L^2}{\kappa_5^2} \partial_z^2 \Phi (z, z_H) \vert_{z = 0}$.

\begin{figure*}[t!] \vspace*{-1mm}
\includegraphics[width=0.49\columnwidth]{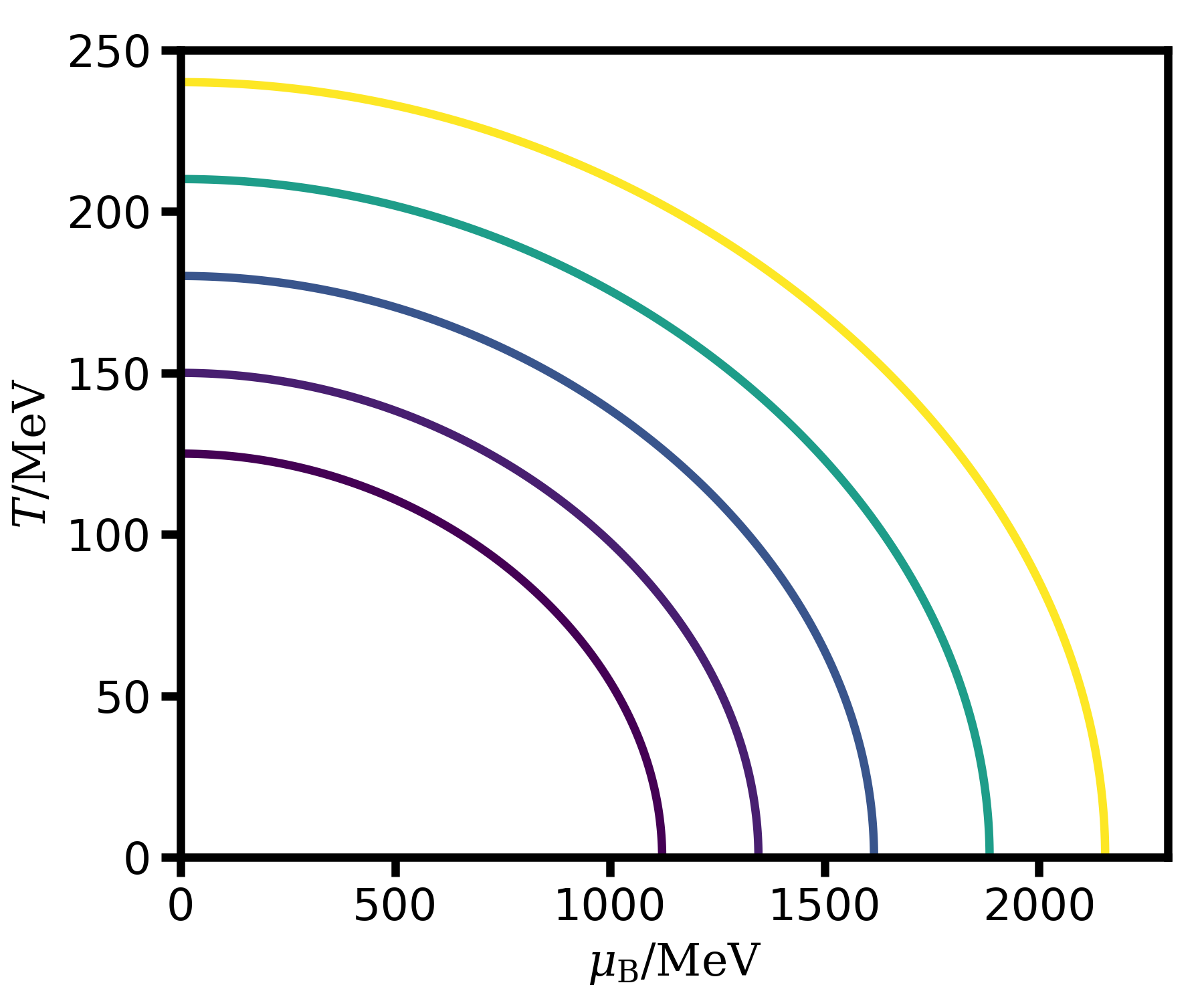} 
\includegraphics[width=0.49\columnwidth]{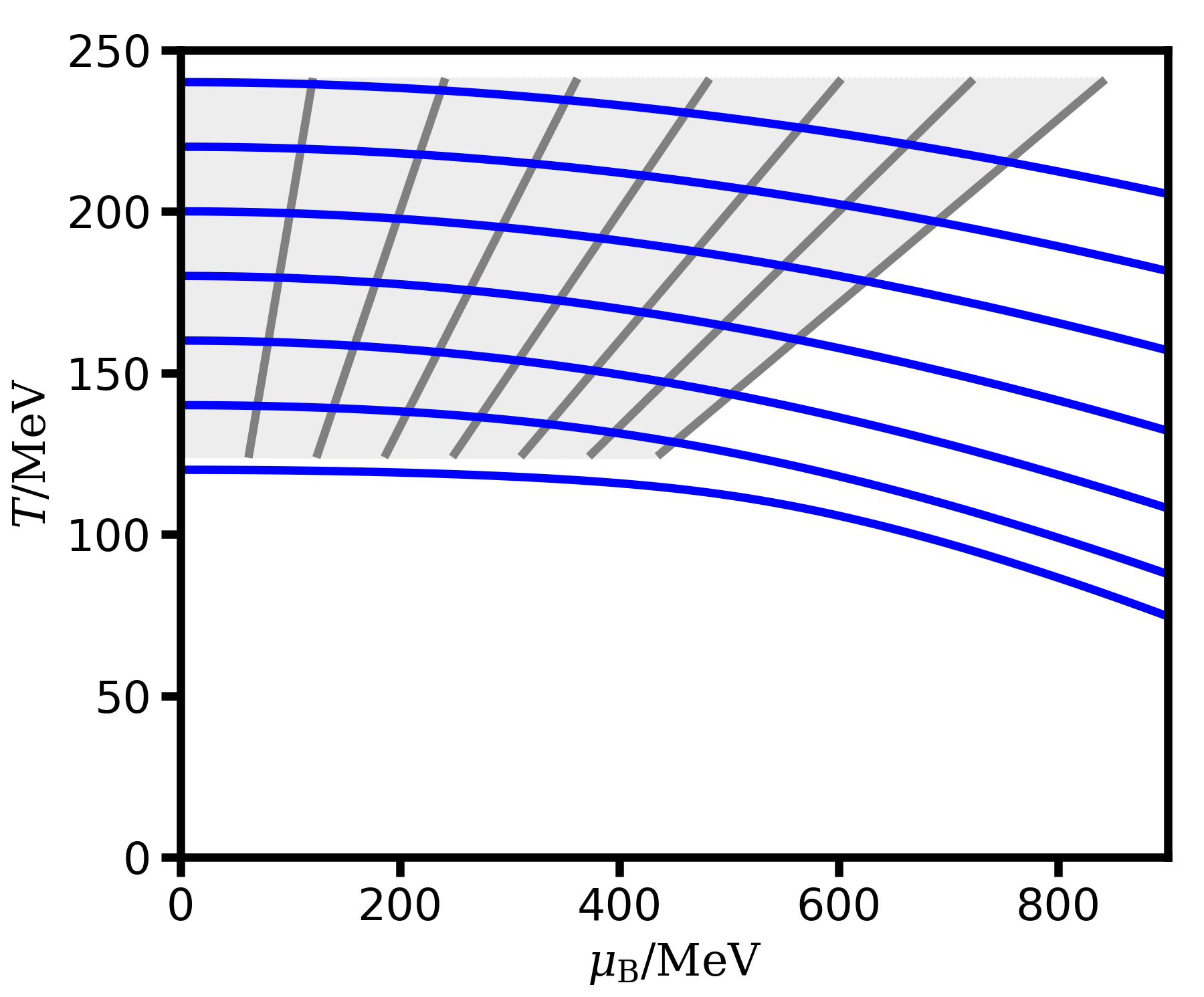}
\caption{Curves of constant pressure over the $T$-$\mu_B$ plane.
In such a way, the pressure data on the $T$ axis are directly ``transported'' towards the $\mu_B$ axis, in particular $p(T=0, \mu_B^{(0)}) = p_0 := p(T_0, \mu_B=0) $ along the constant-pressure curve
$T(\mu_B)\vert_{p=p_0}$ starting at $T(\mu_B = 0) = T_0$ and terminating at $T(\mu_B^{(0)}) = 0$.
The energy density, with $s(T, \mu_B)$ and $n_B (T, \mu_B)$ given, follows then from $e = - p + T s + \mu_B n_B$ (Gibbs-Duhem).
Left panel: A simple toy model is employed here for the purpose of demonstration \blau{($s= 4aT^3 + 2bT \mu_B^2$, $n_B = 4c \mu_B^3 + 2bT^2 \mu_B$, and numerical values $b/a = 0.027384$, $c/a = 0.000154$ referring to a two-flavor ideal quark-gluon plasma)}.
Right panel: For the holographic model~\blau{(\ref{EdM_action}, \ref{eq:dil_pot}, \ref{eq:dyn_coup})} in a region \blau{(grey hatched) controlled by lattice QCD data~\cite{Borsanyi:2021sxv} on the dark-grey beam sections.}
\color{black}
\label{fig:const_p_curves} 
}
\end{figure*}

The comparison of the scaled pressure $p/T^4$, ratio $e/p$ and scaled baryon density $n_B/T^3$ with the data \cite{Borsanyi:2021sxv} at $\mu_B / T = 1$ and 2 in Fig.~\ref{fig:new_hol_plots}
exhibits good agreement in the small-$\mu_B$ region. That is, the model maps successfully QCD thermodynamics from the $T$ axis into the $T$-$\mu_B$ plane,
in particular towards the $\mu_B$ axis:
$p(T) \mapsto p(\mu_B) = p(T(\mu_B))$.
Curves of constant pressure, $T(\mu_B)\vert_{p = const}$, are determined by 
$d T / d \mu_B = - n_B (T, \mu_B) / s (T, \mu_B)$,\footnote{From $p(T(x), \mu_B(x)) = const$ and assuming
a parametric representation $T(x)$ and $\mu_B (x)$ with $x$ being the arc length. 
$d p / d x = (\partial p / \partial T) (d T / d x) + (\partial p / \partial \mu_B) (d \mu_B / d x) = 0$
uses then $s = \partial p / \partial T$ and $n_B = \partial p / \partial \mu_B$. 
The technicalities of determining holographically these quantities and the properties 
of the resulting EoS are relegated to a separate paper. \blau{The left panel in Fig.~\ref{fig:const_p_curves} is for a toy model demonstrating how the EoS given on the temperature axis is mapped to the chemical potential axis.}}
exhibited in Fig.~\ref{fig:const_p_curves}.
The lattice data \cite{Borsanyi:2021sxv} are for $T \in [125, 240]$~MeV at small values of $\mu_B$ \blau{(see right panel)}.
The ``transport'' into the restricted region of the $T$-$\mu_B$ plane by the holographic model 
is already an extrapolation, which requires a smooth pattern of $T(\mu_B)\vert_{p = const}$ curves
without junctions, branchings, crossings etc.
The extrapolation of the data to $T < 125$~MeV and $T > 240$~MeV
at $\mu_B = 0$ by the parameterizations (\ref{eq:dil_pot}, \ref{eq:dyn_coup})
is another type of extrapolation. Thus the region beyond the displayed $p = const$ curves
could be hampered by both types of uncertainties,
but nevertheless may serve as educated guess of the cool EoS in the high-density realm. 
The benefit of such an approach is direct coupling of the hot EoS, being an essential input in
describing ultra-relativistic heavy-ion collisions by fluid dynamics, and the cool EoS,
being the input for compact (neutron) star structure and (merger) dynamics. 

Instead of \blau{explicitly} using hadronic degrees of freedom to devise the holographic EoS, here the EoS as relation of energy density and pressure is solely deployed. Of course, the underlying hot QCD EoS is anchored in the common quark-gluon dynamics with intimate contact to hadron observables. 

\end{document}